\documentclass[a4paper,12pt]{article}
\usepackage{amsfonts}
\usepackage{latexsym}
\usepackage{amsmath}
\usepackage{amssymb}
\usepackage{amssymb}
\usepackage{slashed}
\usepackage{upgreek}
\usepackage{mathrsfs}
\usepackage{bbm}

\usepackage[toc,page]{appendix}

\usepackage{amsthm}
\usepackage[utf8]{inputenc}
\usepackage{graphicx}

\theoremstyle{remark}

\usepackage[utf8]{inputenc}
\usepackage[english]{babel}

\theoremstyle{definition}

\allowdisplaybreaks

\usepackage{relsize}
\usepackage{graphicx}
\usepackage{comment}

\renewcommand{\thefootnote}{\fnsymbol{footnote}}

\def\appendix#1{\addtocounter{section}{1}\setcounter{equation}{0}
	\renewcommand{\thesection}{\Alph{section}}
	\section*{Appendix \thesection\protect\indent \parbox[t]{11.15cm}{#1}}
	\addcontentsline{toc}{section}{Appendix \thesection\ \ \ #1}}

\newcommand{\pp}{=\kern-0.40em{\vert}}

\def\bbe{{\bf{e}}}

\font\mybb=msbm10 at 11pt

\def\bb#1{\hbox{\mybb#1}}

\def\bR {\bb{R}}

\newcommand{\bea}{\begin{eqnarray}}
	\newcommand{\eea}{\end{eqnarray}}

\usepackage[usenames,dvipsnames,svgnames,table]{xcolor}	
\usepackage[normalem]{ulem}				
\usepackage{microtype}
\usepackage[colorlinks, citecolor=blue, linkcolor=blue, urlcolor=Maroon, filecolor=Maroon, linktocpage=true]{hyperref} 

\usepackage{chngcntr}
\counterwithout{equation}{section}

\begin{document}

	\begin{center}
		\vspace*{-1.0cm}
		\begin{flushright}
		\end{flushright}

		
		\vspace{2.0cm} {\Large \bf W-symmetries, anomalies and heterotic backgrounds with   SU holonomy } \\[.2cm]
		
		\vskip 2cm
		L.\, Grimanellis, G.\,  Papadopoulos and  E.\, P\'erez-Bola\~nos
		\\
		\vskip .6cm


		\begin{small}
			\textit{Department of Mathematics
				\\
				King's College London
				\\
				Strand
				\\
				London WC2R 2LS, UK}
			\\*[.3cm]
\texttt{loukas.grimanellis@kcl.ac.uk}
			\\
			\texttt{george.papadopoulos@kcl.ac.uk}
			\\
			\texttt{edgar.perez$\underline{~}$bolanos@kcl.ac.uk}
		\end{small}
		\\*[.6cm]

	\end{center}

	\vskip 2.5 cm

	\begin{abstract}
\noindent
We show that the sigma models with target spaces supersymmetric heterotic backgrounds with  $SU(2)$ and $SU(3)$ holonomy are invariant under a W-symmetry algebra generated by the covariantly constant forms  of these backgrounds. The closure of the W-algebra requires additional generators which we identify. We prove that the chiral anomalies of all these symmetries are consistent at one-loop in perturbation theory. We also demonstrate that these  anomalies cancel at the same loop level either by adding suitable finite local counterterms in the sigma model effective action or by assuming a plausible quantum correction to the transformations.  Such a  correction is  compatible with both  the  cancellation mechanism for spacetime frame rotation and gauge transformation anomalies, and  the correction to the heterotic supergravity up to two loops in the sigma model perturbation theory.

	\end{abstract}

	

	\newpage
	
	\renewcommand{\thefootnote}{\arabic{footnote}}

\section{Introduction}

 It has been known for a sometime that $\hat\nabla$-covariantly constant forms generate  symmetries in 2-dimensional supersymmetric sigma models \cite{odake, dvn, phgpw1, phgpw2} with couplings a metric, $g$, and a Wess-Zumino term, $b$, where $\hat\nabla$ is a metric connection with torsion, $H=db$. The sigma model target spaces considered so far are manifolds $N^n$  that admit such a  connection $\hat\nabla$ whose holonomy is included in the groups $U({n\over2})$, $SU({n\over2})$, $Sp({n\over4})$, $Sp({n\over4})\cdot Sp(1)$, $G_2 (n=7)$ and $\mathrm{Spin}(7) (n=8)$.  As these symmetries arise from the reduction of the holonomy of $\hat\nabla$ to a subgroup of $O(n)$, they  are known {\it holonomy symmetries}. It has been found that the algebra of holonomy symmetries is a W-algebra \cite{phgpw1, phgpw2}, i.e. the structure constants of the algebra depend on the conserved currents of the theory. The structure of these algebras has been explored both in the classical and quantum theory, see  \cite{phgpw1, phgpw2, vafa, jfof, gpph, mg}.
More recently in \cite{dlossa}, some of these backgrounds have been considered as target spaces of heterotic sigma models and the chiral anomalies of these symmetries have been investigated.

Symmetries of heterotic sigma models are  a priori anomalous in the quantum theory because of the presence of chiral worldsheet fermions in the sigma model actions  \cite{gmpn, laggin, bny}.  To preserve the geometric interpretation of these theories, the anomalies of some of these symmetries must cancel. Indeed,  the anomalies  of spacetime frame rotations and the gauge sector transformations cancel after assigning an anomalous variation to the sigma model coupling constant $b$  \cite{ewch}. However, such a variation leads to a non-tensorial transformation law for $H$ which appears as a coupling in many vertices in background field method of quantising the theory.  The restoration of covariance in the quantum theory  requires the modification of $H$ with appropriate Chern-Simons terms \cite{mgjs, sen} at all loops.

The classification of supersymmetric heterotic backgrounds in \cite{uggp1, uggp2} has revealed that there is a much more general class of  spacetimes that can be used as target spaces of heterotic sigma models than those investigated so far in the literature and mentioned above, for a review see \cite{rev}.  These backgrounds   exhibit a variety of $\hat \nabla$-covariantly constant forms constructed as Killing spinor bilinears all of which have been identified. All these forms generate holonomy symmetries in heterotic sigma models.
The purpose of this work is twofold. First,  it is to identify the commutators of all such symmetries, i.e. to identify the algebra of symmetries generated by the covariantly constant form bilinears. Second, it is to find and investigate their chiral anomalies using Wess-Zumino consistency conditions.  We shall demonstrate that the anomalies are consistent at one loop in the sigma model perturbation theory. Moreover, they cancel at the same order  by either adding appropriate finite local counterterms in the sigma model effective action or by assuming that the transformations are appropriately quantum mechanically corrected.

In this paper, we shall focus on two classes of heterotic supersymmetric backgrounds those for which the connection with skew-symmetric torsion, $\hat\nabla$, has holonomy included in $SU(2)$ and in $SU(3)$.  The spacetime, $M^{10}$,  of the former backgrounds can be locally described  as a principal bundle whose fibre is a 6-dimensional  Lorentzian Lie group $G$ with self-dual structure constants and base space a 4-dimensional conformally hyper-K\"ahler  manifold \cite{uggp1, uggp2, rev}.  The holonomy symmetries of sigma models on $SU(2)$ holonomy backgrounds  are generated by 1-forms and 2-forms. The 1-forms are those associated to vector fields generated by the action of $G$ on $M^{10}$ with respect to the spacetime metric $g$. In the $SU(3)$ holonomy case, the spacetime is locally a principal bundle with fibre a 4-dimensional Lorentzian Lie group $G$ and base space a conformally balanced K\"ahler manifold with torsion \cite{uggp1, uggp2, rev}. The holonomy symmetries are generated by four 1-forms which are associated to the vector fields generated by $G$ on $M^{10}$ with respect to $g$ as well as one 2-form, $I$, and two 3-forms, $L_1$ and $L_2$.

We find that the closure of the algebra of holonomy symmetries in both cases requires the inclusion of additional generators.  As the sigma models that we shall consider manifestly exhibit a (1,0) worldsheet supersymmetry, the closure of the algebra of symmetries requires  the inclusion of right-handed worldsheet translations and supersymmetry transformations generated by the right-handed energy momentum tensor, $T$, as well as the symmetries generated by the second Casimir operator of the Lie algebra of $G$.  In addition, the closure of the algebra of symmetries of the sigma model on $SU(3)$ holonomy backgrounds requires the symmetries generated by the conserved current which is the product, $TJ_I$,  of  $T$ with the conserved current, $J_I$, of the symmetry generated by the 2-form $I$.  In both cases, the algebra of symmetries is a W-algebra as the structure constants of the algebra depend on the currents  of the symmetries.

For the analysis of the chiral anomalies of the holonomy symmetries, we assume that there is a regularisation scheme which manifestly quantum mechanically preserves the (1,0) worldsheet supersymmetry of the theory.  This can be justified as the perturbation theory for the model can be done in (1,0) superfields. The anomalous part of the effective action for spacetime frame rotations and gauge transformations has been computed in \cite{zumino,chpt2}.  After perhaps adding appropriate finite local counterterms \cite{chpt2} in the effective action of the theory, the anomaly of spacetime frame rotations and transformations of the gauge sector can be brought into the standard form given here in (\ref{lan}) and (\ref{uan}), respectively. Then, upon using Wess-Zumino consistency conditions, one can show that the anomalies of the holonomy symmetries of the sigma model generated by the
$\hat\nabla$-covariantly constant forms can be expressed in terms of the Chern-Simons form of an appropriate connection as in (\ref{Lanom}) up to possibly spacetime frame rotation and gauge transformation invariant terms, see also \cite{dlossa}. The same applies for the anomalies of the additional  symmetries required for the closure of the algebra of holonomy transformations.

We demonstrate that all these anomalies are consistent at one loop provided that the associated Chern-Simons terms are expressed in terms of the frame connection associated to the connection $\check \nabla$, which has torsion $-H$, and the connection that appears in the classical action of the gauge sector. In fact, the anomalous part of the effective action is naturally expressed in terms of these connections \cite{chpt2}. The cancellation of anomalies proceeds into two different ways. One way involves the assertion that the forms that generate the holonomy symmetries are quantum mechanically corrected such that to the given order in perturbation theory are covariantly constant with respect to a new connection $\hat\nabla^{\hbar}$ with skew symmetric torsion $H^{\hbar}$ which includes the Chern-Simons form.
Such a modification of the torsion is also justified as part of the anomaly cancellation mechanism for the spacetime frame rotation and gauge anomalies of the theory  mentioned above. It is also consistent with the finding that the Killing spinor equations of heterotic supergravity retain their form \cite{roo}  up to and including  two loops in the sigma model perturbation theory provided that the 3-form field strength $H$ is replaced by $H^\hbar$. Such a replacement is a consequence of  the cancellation of gravitational anomalies for heterotic supergravity \cite{grsc}.
For some symmetries, there is an alternative cancellation mechanism based on the existence of finite local counterterms which can be added to the effective action to remove the anomalies. Under certain assumptions, we demonstrate that the anomalies of the symmetries generated by 1-forms and 2-forms can be removed with such finite local counterterms. In the conclusions, we also explore the question whether at higher loops in sigma model perturbation theory some holonomy symmetries receive an anomalous contribution which is invariant under both spacetime frame rotations and gauge sector transformations.

This paper is organised as follows. In section 2, we introduce the action of the heterotic sigma model and present its sigma model and holonomy symmetries. We also give the main formulae for the anomalies, present the consistency conditions and discuss the mechanisms for anomaly cancellation. In section 3, we present the commutators of the symmetries of sigma model with target space the heterotic backgrounds with  $SU(2)$ holonomy. We also give the anomalies of these symmetries, prove that there are consistent and describe their cancellation. In section 4, we describe similar results for the symmetries and anomalies of sigma models with target space heterotic backgrounds with $SU(3)$ holonomy. In section 5, we give our conclusions. In addition in appendix A, we present some key formulae required for the computation of the commutators of holonomy symmetries,  and in appendix B, we give some more details of the computation of the commutators of symmetries of sigma models on backgrounds with $SU(3)$ holonomy.

\section{Classical symmetries of chiral sigma models}

\subsection{Action and sigma model symmetries}

 The classical fields of the (1,0)-supersymmetric 2-dimensional sigma models that we shall investigate  are  maps, $X$, from the worldsheet superspace $\Xi^{2|1}$ with coordinates $(\sigma^=, \sigma^{\pp}, \theta^+)$ into a spacetime $M$, $X:~\Xi^{2|1}\rightarrow M$, and Grassmannian odd sections $\psi$ of a vector bundle $S_-\otimes X^*E$ over $\Xi^{2|1}$, where $S_-$ is the anti-chiral spinor bundle over $\Xi^{2|1}$ and $E$ is a vector bundle over $M$. An action\footnote{The action can include a potential term \cite{chgppt}. But we shall not consider this here.}   for these fields \cite{ewch} is
\bea
S=-i \int d^2\sigma d\theta^+  \Big ((g_{\mu\nu}+b_{\mu\nu}) D_+X^\mu \partial_=X^\nu+i h_{\mathrm{a}\mathrm{b}} \psi_-^\mathrm{a}{\cal D}_+\psi_-^\mathrm{b}\Big)~,
\label{act1}
\eea
where $g$ is a spacetime metric, $b$ is a locally defined 2-form on $M$ such that $H=db$ is a globally defined 3-form,  $D_+^2=i\partial_{\pp}$, $h$ is a fibre metric on $E$ and
\bea
{\cal D}_+\psi_-^{\mathrm{a}}=D_+\psi_-^{\mathrm{a}}+ D_+X^\mu \Omega_\mu{}^{\mathrm{a}}{}_{\mathrm{b}} \psi_-^{\mathrm{b}}~,
\eea
with $\Omega$ a connection on $E$ with curvature $F$. We take without loss of generality that ${\cal D}_\mu h_{\mathrm{a}\mathrm{b}}=0$. We shall refer to the part of the action with couplings $h$ and ${\cal D}$ as the gauge sector of the theory. Note that
\bea
\delta S= -i \int d^2\sigma d\theta^+ \big(\delta X^\mu {\cal S}_\mu+\Delta \psi_-^{\mathrm{a}} {\cal S}_{\mathrm{a}}\big)~,
\eea
where
\bea
\Delta \psi_-^{\mathrm{a}} \equiv \delta \psi_-^{\mathrm{a}} + \delta X^\mu \Omega_\mu{}^{\mathrm{a}}{}_{\mathrm{b}} \psi_-^{\mathrm{b}}~,
\eea
is the covariantisation of $\delta \psi_-$ and
\bea
&&
{\cal S}_\mu=-2 g_{\mu\nu} \hat\nabla_= D_+ X^\nu-i  \psi_-^{\mathrm{a}} \psi_-^{\mathrm{b}} D_+X^\nu F_{\mu\nu \mathrm{a}\mathrm{b}}~,
\cr
&&
{\cal S}^{\mathrm{a}}=2i{\cal D}_+\psi_-^{\mathrm{a}}~,
\label{feqns}
\eea
are the field equations. In addition, the connection, $\hat\nabla$, is
\bea
\hat \nabla_\nu Z^\mu=\nabla_\nu Z^\mu+{1\over2} H^\mu{}_{\nu\lambda} Z^\lambda~,
\eea
where $\nabla$ the Levi-Civita connection on $M$ with respect to the metric $g$ and $H$ is the torsion of $\hat\nabla$ which is skew-symmetric.

{\it Sigma model symmetries} are those transformations  of the fields $X$ and $\psi$ and  the coupling constants $g$, $b$, $h$ and $\Omega$ such that the action (\ref{act1}) remains invariant. This is to distinguish them from the standard symmetries of a field theory which act only on the fields and leave the action invariant.  Clearly such transformations are the diffeomorphisms of the target space $M$ as well as the gauge transformations of the gauge sector. The fields and coupling constants under infinitesimal diffeomorphisms generated by the vector field $v$ transform as
\bea
&&\delta X^\mu=v^\mu~,~~~\delta g_{\mu\nu}=-{\cal L}_v g_{\mu\nu}~,~~~\delta b_{\mu\nu}=-{\cal L}_v b_{\mu\nu}~,
\cr
&&\delta \psi^{\mathrm{a}}_-= w^{\mathrm{a}}{}_{\mathrm{b}} \psi^{\mathrm{b}}_-~,~~~\delta\Omega_\mu{}^{\mathrm{a}}{}_{\mathrm{b}}=-{\cal L}_v\Omega_\mu{}^{\mathrm{a}}{}_{\mathrm{b}}-\partial_\mu w^{\mathrm{a}}{}_{\mathrm{b}}+ w^{\mathrm{a}}{}_{\mathrm{c}}\,  \Omega_\mu{}^{\mathrm{c}}{}_{\mathrm{b}} -\Omega_\mu{}^{\mathrm{a}}{}_{\mathrm{c}}\,  w^{\mathrm{c}}{}_{\mathrm{b}}~,
\cr
&&
\delta h_{\mathrm{a}\mathrm{b}}=-{\cal L}_v h_{\mathrm{a}\mathrm{b}}-  w^{\mathrm{c}}{}_{\mathrm{a}} h_{\mathrm{c}\mathrm{b}}- h_{\mathrm{a}\mathrm{c}}    w^{\mathrm{c}}{}_{\mathrm{b}}~,
\label{diff}
\eea
where it is assumed that the diffeomophisms generated by the vector field $v$ lift to the vector bundle $E$  and generate a fibre rotation\footnote{Typically, one sets $w(v)=0$ as $w(v)$s generate gauge transformations which are independent sigma model symmetries of the system.} $w=w(v)$.  The fields and coupling constants under the infinitesimal gauge transformations of the gauge sector transform as
\bea
&&\delta_u \psi^{\mathrm{a}}_-= u^{\mathrm{a}}{}_{\mathrm{b}}\psi^{\mathrm{b}}_-~,~~~\delta_u\Omega_\mu{}^{\mathrm{a}}{}_{\mathrm{b}}=-\partial_\mu u^{\mathrm{a}}{}_{\mathrm{b}}+ u^{\mathrm{a}}{}_{\mathrm{c}}\,  \Omega_\mu{}^{\mathrm{c}}{}_{\mathrm{b}} -\Omega_\mu{}^{\mathrm{a}}{}_{\mathrm{c}}\,  u^{\mathrm{c}}{}_{\mathrm{b}}~,
\cr
&&\delta_u h_{\mathrm{a}\mathrm{b}}=-  u^{\mathrm{c}}{}_{\mathrm{a}} h_{\mathrm{c}\mathrm{b}}- h_{\mathrm{a}\mathrm{c}}    u^{\mathrm{c}}{}_{\mathrm{b}}~,
\label{ftran1}
\eea
where $u$ is the infinitesimal parameter, and the remaining fields and couplings of the theory remain inert.

It is convenient in quantum theory to introduce a frame on the tangent bundle of the spacetime.  This is because in the background field method of computing the
effective action, it is convenient to express the quantum field in a frame basis, see section \ref{sec:anom} for more details. In such a case, if we write the metric as $g_{\mu\nu}= \eta_{AB} \bbe^A_\mu \bbe^B_\nu$, then the action of infinitesimal spacetime frame rotations will be
\bea
\delta_\ell \bbe^A_\mu= \ell^A{}_B  \bbe^B_\mu~,~~~~\delta_\ell \omega_\mu{}^A{}_B=-\partial_\mu \ell^A{}_B+ \ell^A{}_C\,  \omega_\mu{}^C{}_B -\omega_\mu{}^A{}_C\,  \ell^C{}_B~,
\label{ftran2}
\eea
where $\ell$ is the infinitesimal parameter and  $\omega$ is a frame connection of the tangent bundle which we shall always assume that preserves the spacetime metric. Of course $\omega$ transforms under diffeomorphisms as $\Omega$ in (\ref{diff}) as well as the spacetime (co-)frame $\bbe^A$.

 The commutator of two  gauge transformations (\ref{ftran1}) is $[\delta_u, \delta_{u'}]=\delta_{[u, u']}$, where $[\cdot, \cdot]$ is the usual commutator of two matrices.
 A similar result holds for the commutator of two spacetime frame rotations (\ref{ftran2}). In addition to these, there is a gauge symmetry
\bea
\delta b_{\mu\nu}=(dm)_{\mu\nu}~,
\label{gbtrans}
\eea
associated with 2-form gauge potential, where $m$ is a 1-form on the spacetime.

\subsection{Holonomy symmetries}\label{sec:g}

Let $L$  be a vector $\ell$-form $L$ on the sigma model target space $M$ and consider the infinitesimal transformation
\bea
\delta_L X^\mu=a_L L^\mu{}_{\lambda_1\dots \lambda_\ell} D_+X^{\lambda_1}\dots D_+X^{\lambda_\ell}\equiv a_L L^\mu{}_L D_+X^L~,~~~\Delta_L\psi_-^\mathrm{a}=0~,
\label{Gsym}
\eea
where $a_L$ is the parameter chosen such that $\delta_L X^\mu$ is even under Grassmannian parity, the index $L$ is the multi-index $L=\lambda_1 \dots \lambda_\ell$ and $D_+X^L=D_+X^{\lambda_1}\cdots D_+X^{\lambda_\ell}$.  Such transformations \cite{phgpw1, phgpw2} leave the action  (\ref{act1}) invariant provided that, after lowering the vector index with the metric $g$, $L$ is an $(\ell+1)$-form and
\bea
\hat\nabla_{\nu} L_{\lambda_1\dots\lambda_{\ell+1}}=0~,~~~F_{\nu[\lambda_1} L^\nu{}_{\lambda_2\dots \lambda_{\ell+1}]}=0~.
\label{invcon}
\eea
The second condition\footnote{This can also be written as $i_LF=0$, where $i_L$ is the inner derivation of $F$ with respect to the vector $\ell$-form $L$. In general, the inner derivation of a p-form, $P$, with respect to $L$ is $i_LP={1\over \ell!(p-1)!}  L^\nu{}_L P_{\nu P_2} dx^{LP_2}$, where $P_2=\mu_2\dots \mu_{p}$.} above has appeared in \cite{dlossa} and generalises the condition required for the model to admit (2,0) worldsheet supersymmetry, see \cite{chgppt}.
 Moreover, the parameter $a_L$ satisfies $\partial_=a_L=0$, i.e. $a_L=a_L(x^{\pp}, \theta^+)$. It is straightforward to observe that for $\ell=0$, $L=K$ is a Killing vector field and $i_K H=dK$. As $H$ is a closed 3-form, ${\cal L}_K H=0$.

 The existence of $\hat\nabla$-covariantly constant forms implies the reduction\footnote{The structure group of the spacetime reduces as well. The structure group is a subgroup of the holonomy group of any connection.} of the holonomy group of $\hat\nabla$  to a proper subgroup of $SO$. This occurs in a number of scenarios that include string compactifications on special special holonomy manifolds.
 More recently it has been demonstrated in \cite{uggp1, uggp2} that the geometry of all supersymmetric heterotic backgrounds is characterised by the reduction of the holonomy group of the $\hat\nabla$ connection to a subgroup of $SO(9,1)$. This reduction of the holonomy group of the $\hat\nabla$ connection  leads to the existence of $\hat\nabla$-covariantly constant forms on the spacetime which are constructed as Killing spinor bilinears. In addition, the second condition in (\ref{invcon}) is a consequence of the gaugino Killing spinor equation. Therefore, all these covariantly constant forms generate symmetries of the action (\ref {act1}) which will be investigated below.

The commutator of two transformations (\ref{Gsym}) on the field $X$ has been explored in detail in  \cite{gpph}. Here, we shall summarise some of the formulae and emphasise some differences as some assumptions made previously on the algebraic properties of the $\hat\nabla$-covariantly constant tensors are not valid for those of general supersymmetric heterotic backgrounds. Moreover, we shall describe the commutator of the transformations on the field $\psi$.  The commutator of two  transformations (\ref{Gsym}) on the field $X$ generated by the vector $\ell$-form $L$ and the vector $m$-form $M$  can be written as
\bea
[\delta_L, \delta_M]X^\mu= \delta_{LM}^{(1)} X^\mu+\delta_{LM}^{(2)} X^\mu+\delta_{LM}^{(3)} X^\mu~,
\label{comm}
\eea
with
\bea
\delta_{LM}^{(1)} X^\mu=a_M a_L N(L,M)^\mu{}_{LM} D_+X^{LM}~,
\eea
\bea
(\delta_{LM}^{(2)} X)_\mu&=&\big (-m a_M D_+a_L (L\cdot M)_{\nu L_2, \mu M_2}
\cr
&&
\qquad\qquad
+ \ell (-1)^{(\ell+1) (m+1)} a_L D_+ a_M (L\cdot M)_{\mu L_2,\nu M_2}\big)
 D_+X^{\nu L_2M_2}~,
\eea
and
\bea
(\delta_{LM}^{(3)} X)_\mu=-2i \ell m (-1)^\ell a_M a_L (L\cdot M)_{(\mu|L_2|, \nu )M_2} \partial_{\pp}X^\nu D_+X^{L_2M_2}~,
\eea
where
\bea
(L\cdot M)_{\lambda L_2,\mu M_2}=L_{\rho \lambda[L_2} M^\rho{}_{|\mu| M_2]}~,
\eea
and
\bea
N(L,M)^\mu{}_{LM} dx^{LM}&=&\Big (L^\nu{}_L\partial_\nu M^\mu{}_{M}- M^\nu{}_M \partial_\nu L^\mu{}_L-\ell L^\mu{}_{\nu L_2} \partial_{\lambda_1} M^\nu{}_M
\cr
&&
+m M^\mu{}_{\nu M_2}
\partial_{\mu_1} L^\nu{}_L\Big) dx^{LM}~.
\eea
The multi-index $M$ stands for $M=\mu_1\dots\mu_m$ while the multi-indices $L_2$ and $M_2$ stand for $L_2=\lambda_2\dots\lambda_\ell$ and $M_2=\mu_2\dots \mu_m$, respectively.
The tensor $N(L,M)$ is the Nijenhuis tensor of the vector-forms $L$ and $M$ and reduces to the standard Nijenhuis tensor for an almost complex structure $I$ upon setting in the above expression $L=M=I$. Furthermore after using that $L$ and $M$ are $\hat\nabla$-covariantly constant, the Nijenhuis tensor can be rewritten as
\bea
&&N_{\mu LM} dx^{LM}=\Big(-H_{\mu\nu\rho} L^\nu{}_L M^\rho{}_M+mL^\nu{}_L H_{\nu\mu_1}{}^\rho M_{\mu\rho M_2}-\ell M^\nu{}_MH_{\nu\lambda_1}{}^\rho L_{\mu\rho L_2}
\cr
&&\qquad\qquad\qquad
+\ell m H^{\rho}{}_{\lambda_1\mu_1} ( L\cdot M)_{(\mu|L_2|, \rho)M_2}\Big)\, dx^{LM}
\cr
&&\qquad=  \Big(- (\ell+m+1)H_{[\mu|\nu\rho|} L^\nu{}_L M^\rho{}_{M]}+\ell m H^{\rho}{}_{\lambda_1\mu_1} ( L\cdot M)_{(\mu|L_2|, \rho)M_2}\Big)\, dx^{LM} ~.
\eea
This concludes the description of the commutator of two holonomy symmetries. The conserved current of a symmetry generated by the $(\ell+1)$-form $L$ is
\bea
J_L=L_{\mu_1\dots\mu_{\ell+1}} D_+X^{\mu_1\dots\mu_{\ell+1}}~.
\label{curL}
\eea
In can be easily seen that $\partial_=J_L=0$ subject to field equations (\ref{feqns}).

To investigate the commutator (\ref{comm}), we have to also consider symmetries generated by  the vector (q+1)-form
\bea
S^\mu{}_{\nu\rho_1\dots \rho_q}= g^{\mu\lambda} S_{\lambda, \nu\rho_1\dots \rho_q}= \delta^\mu{}_{[\nu} Q_{\rho_1\dots \rho_q]}~.
\label{strans1}
\eea
It turns out that if $Q$ is a $\hat\nabla$-parallel q-form and $i_Q F=0$, i.e. it satisfies (\ref{invcon}),  one can show that the infinitesimal transformation
\bea
\delta_S X_\mu&=&\alpha_S \hat\nabla_+ D_+X^\nu S_{\nu,\mu Q} D_+X^Q+{(-1)^q\over q+1} \hat\nabla_+(\alpha_S S_{\mu,\nu Q} D_+X^{\nu Q})
\cr
&-& \quad{q+3\over3 (q+1)} \alpha_S H_{[\mu\nu\rho}
Q_{Q]} D_+X^{\mu\nu\rho Q}~,~~~
\cr
\Delta\psi^{\mathrm a}_-&=&-{(-1)^q\over q+1} \alpha_S Q_Q F_{\mu\nu}{}^{\mathrm a}{}_{\mathrm b} \psi_-^{\mathrm b} D_+X^{Q\mu\nu}~,
\label{strans2}
\eea
is a symmetry of the action.  It will be significant later for the analysis of the anomaly consistency conditions  that the proof of invariance of the action (\ref{act1}) under the transformations  (\ref{strans1})  requires the Bianchi identity
\bea
\hat R_{\mu [\nu,\rho\sigma]}=-{1\over3} \hat\nabla_\mu H_{\nu\rho\sigma}~,
\label{bia1}
\eea
where $H$ is taken to be a closed form, $dH=0$.
The associated conserved current is $T J_Q$, where $T$ is the right-handed (super) energy-momentum tensor
\bea
T_{+\pp}=g_{\mu\nu} D_+X^\mu \hat\nabla_+ D_+ X^\nu-{1\over3} H_{\mu\nu\rho} D_+X^{\mu\nu\rho}~,
\eea
which does not depend on the left-handed fermionic superfield $\psi$.
It can be easily demonstrated using (\ref{bia1})
that $\partial_= T=0$ subject to the field equations (\ref{feqns}).
Note that the infinitesimal transformation generated by $T$ is
\bea
\delta_T X^\mu=2i\alpha_{{}_T} \partial_{\pp} X^\mu+D_+\alpha_{{}_T} D_+X^\mu~,~~~\Delta_T\psi^{\mathrm a}_-=-\alpha_{{}_T} F_{\mu\nu}{}^{\mathrm a}{}_{\mathrm b} \psi_-^{\mathrm b} D_+X^{\mu\nu}~,
\label{1susy}
\eea
which is a right-handed worldsheet translation followed by a right-handed supersymmetry transformation. The prove that the action (\ref{act1}) is invariant under (\ref{1susy})
again requires the use of the Bianchi identity (\ref{bia1}).

To proceed, the commutator (\ref{comm}) has to be re-organised as a sum of variations with each variation  be  independently a symmetry of the action. Such a   re-organisation has first been suggested for the models with $H=0$ in \cite{sven} and later in the models with $H\not=0$  in   \cite{gpph}. However, the conditions on the $\hat\nabla$-covariantly constant forms mentioned in these references necessary for this to work  are not met by the $\hat\nabla$-covariantly constant forms of general supersymmetric heterotic backgrounds that we shall be considering. Locally, all supersymmetric heterotic backgrounds with compact holonomy group are  fibrations with fibre a group manifold \cite{uggp1, uggp2, rev}. The conditions mentioned in \cite{sven, gpph} fail along the fibre directions but still apply provided that the right-hand-side of the commutator (\ref{comm}) is restricted along orthogonal directions  to those of the fibres. Because of this, the formulae in \cite{gpph} are still useful for the analysis and they are summarised in appendix A. Eventually, the whole commutator (\ref{comm}) of holonomy symmetries generated by the $\hat\nabla$-covariantly constant forms of general supersymmetric heterotic backgrounds  can be rewritten as a sum of symmetries. But as we shall demonstrate, this will require the addition of new generators which will be investigated on a case by case basis.

Next, let us consider the commutator of two (\ref{Gsym}) transformations on the field $\psi$. As $\Delta_L\psi=\Delta_M\psi=0$, one finds that
\bea
[\delta_L, \delta_M]\psi^{\mathrm a}_-=-\Omega_\mu{}^{\mathrm a}{}_{\mathrm b} [\delta_L, \delta_M]X^\mu \psi^{\mathrm b}-F_{\mu\nu}{}^{\mathrm a}{}_{\mathrm b} \delta_L X^\mu \delta_M X^\nu \psi^{\mathrm b}_-~.
\label{psicom}
\eea
Therefore, the commutator may give rise to a non-trivial transformation, $\Delta_{LM}$, on $\psi$ given by
\bea
\Delta_{LM}\psi^{\mathrm a}_-=-F_{\mu\nu}{}^{\mathrm a}{}_{\mathrm b} \delta_L X^\mu \delta_M X^\nu \psi^{\mathrm b}_-~.
\eea
Such a transformation may have been expected for consistency as the invariance of the action under (\ref{1susy}) and (\ref{strans2}), which appear in the right hand side of the commutator of two holonomy transformations,  require such a contribution.

For the analysis of anomalies described below, one needs to find the commutators of sigma model and holonomy symmetries. As it can always be arranged for the diffeomorphism sigma model symmetries not to be anomalous up to possibly the addition of a finite local counterterm in the effective action \cite{zumino}, it remains to describe the commutator of gauge transformations (\ref{ftran1}) and (\ref{ftran2}) with the holonomy symmetries (\ref{Gsym}). It is straightforward to see that they commute
\bea
[\delta_\ell, \delta_L]=[\delta_u, \delta_L]=0~,
\label{comgG}
\eea
on both $X$ and $\psi$ fields.

\subsection{Anomaly consistency  conditions}\label{sec:anom}

To retain most of the geometric properties of sigma models in the quantum theory, one uses the background field method \cite{honer, friedan, mukhi} to quantise the theory. This involves the splitting of the (total) field of the theory into a background field, that it is treated classically, and a quantum field that it is quantised. However, for sigma models this splitting is non-linear and because of this the understanding of the quantum theory  presents several challenges. These include the non-linear split symmetry \cite{phgpks} which is required to control the counterterms in order to correctly subtract the ultraviolet infinities and determine the effective action, $\Gamma$, from the 1PI diagrams with only external background lines\footnote{Potential anomalies in the shift symmetry have been examined in \cite{blasi}.}.  After writing the theory in terms of background and quantum fields and considering  the effective action constructed from 1PI diagrams with external background lines only, some of the symmetries of the theory, like those of spacetime frame rotations and gauge sector transformations, act linearly on the quantum fields, for a detailed discussion see \cite{phgpa}. Such symmetries are much more straightforward to investigate. However, this is not the case for the holonomy symmetries, where the induced transformations on the quantum fields is non-linear, and a much more in depth analysis is required \cite{sven}. To proceed, we shall consider the effective action, $\Gamma$, computed from 1PI diagrams with only external background lines.  Then, after stating the spacetime frame rotations and gauge transformation anomalies, which we take them to be expressed in terms of the background fields, we shall use Wess-Zumino consistency conditions to determine the anomalies of the holonomy symmetries. From now on, it will be assumed  that all fields that enter in the expression for the anomalies as well as those that appear in the various transformations  required for the investigation are the background fields.

Suppose that the classical theory
is invariant under the algebra of symmetries whose variations on the fields satisfy the commutation relations
\bea
[\delta_A, \delta_B]=\delta_{[A,B]}~,
\label{cccom}
\eea
where $\delta_A$ ($\delta_B$) is a transformation on the (background) fields generated by $A$   ($B$) generator with parameter  $a_{{}_A}$ ($a_{{}_B}$) and $[A,B]$ is the commutator of the two generators. If these symmetries are anomalous in the quantum theory, i.e. $\delta_A \Gamma= \Delta(a_{{}_A})$,
 then  applying the commutator (\ref{cccom}) on $\Gamma$, one finds that
\bea
\delta_A\Delta(a_{{}_B})-\delta_B\Delta(a_{{}_A})=\Delta(a_{{}_{[A,B]}})~.
\label{WZcon}
\eea
These relations between anomalies are known as Wess-Zumino anomaly consistency conditions\footnote{It is customary in the investigation of Wess-Zumino consistency conditions for anomalies to use the BRST formalism. We shall not do this here.  Instead, we shall use the commutators as these emphasise the geometry structure of the theory.}. A solution of these conditions will yield an expression for the anomaly of a symmetry in terms of the fields.

It is well known that the anomaly associated to the gauge transformations (\ref{ftran2}) is determined by the descent equations \cite{zumino} starting from a 4-form, $P_4(R)= \mathrm{tr} (R(\omega)\wedge R(\omega))$, which is proportional to the first Pontryagin
form of the manifold, where $R$ is the curvature of a connection $\omega$. As this is closed,  one can locally write $P_4(R)=dQ^0_3(\omega)$, where $Q^0_3$ is the Chern-Simon form. As $P_4$ is invariant under the gauge transformations (\ref{ftran2}), one has that $d\delta_\ell Q^0_3(\omega)=0$ and so $\delta_\ell Q^0_3(\omega)=dQ^1_2(\ell, \omega)$. The gauge anomaly\footnote{For applications to string theory, replace in the formulae below $\hbar$ with $\alpha'$.} is given by
\bea
\Delta(\ell)={i\hbar\over 4\pi} \int d^2\sigma d\theta^+ Q^1_2(\omega, \ell)_{\mu\nu} D_+X^\mu \partial_=X^\nu~,
\label{lan}
\eea
where the numerical coefficient in front is determined after an explicit computation of the term in the effective action that contributes to the anomaly. A similar calculation reveals that the anomaly of the gauge transformation (\ref{ftran1}) is
\bea
\Delta(u)=-{i\hbar\over 4\pi} \int d^2\sigma d\theta^+ Q^1_2(\Omega, u)_{\mu\nu} D_+X^\mu \partial_=X^\nu~.
\label{uan}
\eea
The connection that appears in the expressions for the anomalies can be altered upon adding a finite local counterterm in the effective action and therefore it is not uniquely defined \cite{zumino}. This freedom of choosing the connection that appears in the expressions for the anomaly will used later to demonstrate
the consistency of Wess-Zumino conditions. Furthermore, notice that although there is a standard expression for the Chern-Simons form,  $Q^0_3$ is specified up to an exact form, $Q^0_3\rightarrow Q^0_3+dW$. It turn out that this can be used to cancel some of the anomalies with the addition of  an appropriate finite local counterterm constructed from $W$ in the effective action.

As the  commutator of gauge  symmetries with the holonomy symmetries vanishes (\ref{comgG}), the  anomaly consistency conditions (\ref{WZcon}) in this case imply that
\bea
\delta_\ell \Delta(a_{{}_L})-\delta_L \Delta(\ell)=0~,
\eea
and similarly for the gauge transformations (\ref{ftran1}), where $ \Delta(a_{{}_L})$ is the anomaly of the holonomy symmetry generated by $L$. A solution to both consistency conditions is
\bea
\Delta(a_{{}_L})={i\hbar\over 4\pi} \int d^2\sigma d\theta^+\, Q_3^0(\omega, \Omega)_{\mu\nu\rho} \delta_L X^\mu D_+X^\nu \partial_=X^\rho+ \Delta_{\mathrm {inv}}(a_{{}_L})~,
\label{Lanom}
\eea
up to possibly $\delta_\ell$- and $\delta_u$-invariant terms, $\Delta_{\mathrm inv}(a_{{}_L})$, where $Q_3^0(\omega, \Omega)=Q_3^0(\omega)- Q_3^0(\Omega)$. This form of the holonomy symmetry anomaly is also consistent with the second commutator in (\ref{comgG}). From here on, we shall take $\Delta_{\mathrm inv}(a_{{}_L})=0$. Though, we shall comment on the  existence of such a contribution in the anomaly later on in the conclusions.

In the case that $L$ is a Killing vector field $K$, $L=K$ such that $i_K P_4=0$, one has that ${\cal L}_KP_4=0$ and so $d{\cal L}_K Q_3^0=0$.  Thus ${\cal L}_K Q_3^0(\omega)=dQ^1(a_{{}_K}, \omega)$. But ${\cal L}_K Q_3^0(\omega)=di_K Q^0_3(\omega)+i_L P_4(R)=dQ^1_2(a_K, \omega)$.  Thus $i_K Q^0_3(\omega)=Q^1_2(a_K, \omega)$ up to a closed two form which can be absorbed in the definition of $Q^1_2(a_K, \omega)$. Therefore, the anomaly in this case can be written as
\bea
\Delta(a_K)={i\hbar\over 4\pi} \int d^2\sigma d\theta^+ Q^1_2(\omega, a_K)_{\mu\nu} D_+X^\mu \partial_=X^\nu~.
\eea
 $\Delta(a_K)$ satisfies the consistency conditions that arise from  the commutator of isometries of the  sigma model target space.

It remains to investigate the consistency of (\ref{Lanom}) with respect to the commutators of holonomy symmetries. Consider two holonomy symmetries generated by the forms $L$ and $M$. After a direct computation, one finds that
\bea
&&\delta_L \Delta(a_{{}_M})-\delta_M \Delta(a_{{}_L})={i\hbar\over 4\pi} \int d^2\sigma d\theta^+ P_4(R, F)_{\mu\nu\rho\sigma} \delta_L X^\mu \delta_M X^\nu D_+ X^\rho
\partial_=X^\sigma
\cr
&&\qquad\qquad
+ {i\hbar\over 4\pi} \int d^2\sigma d\theta^+ Q_3^0(\omega, \Omega)_{\mu\nu\rho} [\delta_L, \delta_M] X^\mu  D_+ X^\nu \partial_=X^\rho~,
\label{LMcon}
\eea
 where $P_4(R, F)=P_4(R)-P_4(F)$.  It turns out that the above consistency condition is more general. If the anomaly of two transformations $\delta_1$ and $\delta_2$ is given as in (\ref{Lanom}), then their mutual consistency condition will be given as in (\ref{LMcon}) with $\delta_L=\delta_1$ and $\delta_M=\delta_2$.

A comparison of (\ref{LMcon}) with the anomaly consistency condition (\ref{WZcon}) reveals that there may be a potential inconsistency due to the terms that contains $P_4(R, F)$ in the right hand side of (\ref{LMcon}). In fact the consistency condition is  more subtle as it also depends on  whether the individual symmetries that appear in the right hand side  of the commutator $[\delta_L, \delta_M]$ are anomalous. If they are not anomalous, then the whole right hand side of (\ref{LMcon}) must vanish for consistency.
 The investigation of these consistency conditions clearly depends  on the symmetries that arise in the commutator $[\delta_L, \delta_M]$.  This in turn depends on the details of geometry of the heterotic backgrounds and in particular of the properties of $L$ and $M$ forms.

 \subsection{Anomaly cancellation and consistency conditions revisited} \label{s:anom}

 It is widely believed that the holonomy  of $\hat\nabla$ for heterotic backgrounds, and so the number of Killing spinors, is preserved in some form under quantum corrections to possibly all orders in perturbation theory. There are two scenarios of how this can happen. To describe these let us focus on the anomaly cancellation at one loop. We shall comment in the conclusions about anomaly cancellation in higher orders. First, one may expect that the anomalies can be removed by finite local counterterms.  This is mostly the case whenever there is a renormalisation scheme such  that $L$, which generates the symmetry, does not receive quantum corrections.  An example of this is to consider a spacetime $\bR^k\times N^{10-k}$ and $L=I$ a complex structure on $N^{10-k}$. In such a case $I$ generates a second supersymmetry for the part of the sigma model action on $N^{10-k}$ and the theory is (2,0) supersymmetric. In complex coordinates that $I$ is constant, the (2,0) supersymmetry transformations are linear in the fields.  So one does not expect these transformations  to be corrected in the quantum theory. Moreover, one does not expect a (2,0) supersymmetry anomaly because the perturbation theory can be set up using (2,0) superfields.  These  manifestly preserve the symmetry. Therefore, there must be a renormalisation scheme that manifestly preserves the (2,0) supersymmetry. Indeed, it has been shown  that if the perturbation theory is set up in (1,0) superfields, then there is a finite local counterterm \cite{sen, phgpa} which cancels the anomaly (\ref{Lanom}) of the symmetry generated by $I$.

 Second, a plausive scenario is that $L$ will receive quantum corrections.  Such a scenario is consistent with the fact that the Killing spinor equations of heterotic supergravity retain their form  up to and including two loops\footnote{Therefore from the supergravity perspective, the anomalies cancel up to and including two loops in the sigma model perturbation theory.} in the sigma model perturbation theory \cite{roo}. The only modification needed is to replace $H$ with $H^\hbar$ as it will be explained below. Indeed, let us denote the quantum corrected $L$ with $L^{\hbar}$.  To specify the nature of these quantum corrections, one observes that (\ref{Lanom}) in terms of $L^{\hbar}$ can be rewritten as
 \bea
 &&\delta^\hbar_{L}\Gamma=\delta^\hbar_{L}(\Gamma^{(0)}+\hbar \Gamma^{(1)})=\Delta_L(a_L)\Longrightarrow
  \cr
  &&\qquad-i \int d^2\sigma d\theta^+  \big(a_L {2(-1)^\ell\over \ell+1}\hat\nabla^{\hbar}_\mu L^{\hbar}_{L+1} \partial_=X^\mu D_+X^{L+1}-i a_L L^{\hbar}{}^\mu{}_L F^\hbar_{\mu\nu \mathrm{a}\mathrm{b}} \psi^\mathrm{a} \psi^\mathrm{b}  D_+X^{L\nu}
  \cr
  && \qquad+2i \Delta^\hbar_L\psi_-^{\mathrm a} {\cal D}^\hbar_+\psi_{-{\mathrm a}} \big) =0+{\cal O}(\hbar^2)~,
 \label{corran}
 \eea
where $\hat\nabla^{\hbar}$  is the quantum corrected connection\footnote{It is expected that both  $g$ and $b$ are also corrected in quantum theory to $g^\hbar$ and $b^\hbar$, respectively. This is especially the case whenever one searches for a scheme to make the theory manifestly superconformal, i.e. a scheme that the beta function vanishes. So $\hat\nabla^{\hbar}$ should be taken with respect to $g^\hbar$ and $b^\hbar$. But for simplicity in what follows, we shall drop the $\hbar$ superscript from $g$ and $b$. }   with skew-symmetric torsion
\bea
H^{\hbar}=H-{\hbar\over 4\pi} Q^0_3(\omega, \Omega)+{\cal O}(\hbar^2)~.
\label{corrh}
\eea
  Similarly,  ${\cal D}^\hbar$ is the quantum corrected connection of the gauge sector and for the holonomy symmetries $\Delta^\hbar_L\psi=0$.  Nevertheless, as we shall see later this term has to be added as it contributes to the commutator of two holonomy symmetries. Therefore, $d H^{\hbar}=-{\hbar\over4\pi} P_4(\omega, \Omega)$ and it is not closed. Clearly,  the anomaly is cancelled, $\delta^\hbar_{L}\Gamma=0+{\cal O}(\hbar^2)$, provided that $L^\hbar$ is covariantly constant with respect to $\hat\nabla^\hbar$, $\hat\nabla^\hbar L^\hbar=0$,  and $i_{L^\hbar} F^\hbar=0$, i.e.
   the second condition in (\ref{invcon}) is satisfied with $F=F^\hbar$ and  $L=L^{\hbar}$.

Note that the correction  of $H$ as in  (\ref{corrh}) is also required to restore the tensorial properties of the 3-form coupling $H$.  As  the cancellation of the frame rotations and gauge anomalies  assigns an anomalous variation, $\delta_\ell b={\hbar\over 4\pi} Q_2^1(\ell, \omega)$ and $\delta_u b=-{\hbar\over4\pi} Q_2^1(u, \Omega)$, to $b$ at one loop \cite{ewch}. This gives rise to a non-trivial transformation on $H$ at the same loop order which is cancelled in $H^{\hbar}$ by the zeroth order variation of the Chern-Simons term. So  $H^{\hbar}$ is invariant under such a transformation up to order ${\cal O}(\hbar^2)$. This appropriately persists to all loop orders \cite{sen, phgpa}.

The second scenario described above for cancelling holonomy anomalies is also consistent with the corrections to the heterotic supergravity up and including
two loops in the sigma model perturbation theory.  It is known that the Killing spinor equations of the theory remain unaltered to this loop order provided one replaces the 3-form field strength $H$ with $H^\hbar$ \cite{roo}. In turn, this implies that the Killing spinors $\epsilon$ are parallel with respect to the spin connection of $\hat\nabla^\hbar$, $\hat\nabla^\hbar\epsilon=0$. As a consequence the form Killing spinor bilinears can automatically be identified with $L^\hbar$ and they are covariantly constant with respect to $\hat\nabla^\hbar$. Finally, the gaugino Killing spinor equation of the theory implies that $i_{L^\hbar} F^\hbar=0$.

To continue, let us revisit the  consistency condition  (\ref{LMcon}) in view of corrections on $L$ and $M$.  The consistency condition (\ref{LMcon}) can be re-derived  by varying (\ref{corran}) with $\delta^\hbar_{M}$ and taking the commutator. After assuming that $i_{L^\hbar} F^\hbar=i_{M^\hbar} F^\hbar=0$, the final expression can be cast into the form
\bea
&&-i\int d^2\sigma d\theta^+ \big(-2 g_{\mu\nu} [\delta^\hbar_L, \delta^\hbar_M] X^\mu \hat\nabla^{\hbar}_= D_+ X^\nu+ d H^{\hbar}_{\mu\nu\rho\sigma} \delta^{\hbar}_L X^\mu \delta^{\hbar}_M X^\nu D_+X^\rho \partial_=X^\sigma
\cr
&&
+2i (\Delta_{LM}^\hbar\psi^{\mathrm a}_-+F^\hbar_{\mu\nu}{}^{\mathrm a}{}_{\mathrm b} \psi_-^{\mathrm b} \delta_L^\hbar X^\mu \delta_M^\hbar X^\nu) {\cal D}^\hbar_+\psi_{-{\mathrm a}}
\big)=0+{\cal O}(\hbar^2)~,
\label{concon}
\eea
where $\Delta_{LM}^\hbar\psi^{\mathrm a}_-\equiv \delta_{LM}^\hbar\psi^{\mathrm a}_-+\Omega^\hbar_\mu{}^{\mathrm a}{}_{\mathrm b} \delta_{LM}^\hbar X^\mu \psi_-^{\mathrm b}$ and it should be read as the covariantisation  of the right hand side of the commutator on $\psi$ (\ref{psicom})  after it has been decomposed as a sum of individual symmetries.

Further progress depends on the details of the symmetries appearing in the right hand side of the commutator $[\delta^\hbar_L, \delta^\hbar_M]$. It turns out that in all cases the term in (\ref{concon})  involving variations on the field $\psi$ is always satisfied provided that $i_{L^\hbar} F^\hbar=i_{M^\hbar} F^\hbar=0$ that we have already assumed.  In all examples we shall be considering, the commutator on $X$ will read
\bea
[\delta^\hbar_L, \delta^\hbar_M]= \delta^{\hbar}_{N}+\delta^{\hbar}_S+ \delta^\hbar_{J P}~,
\label{comcomp}
\eea
where $\delta^{\hbar}_{N}$ is a symmetry generated by a $\hat\nabla^{\hbar}$-covariantly constant form $N$ with parameter $a_{N}$ constructed
from those of $\delta^{\hbar}_L$ and $\delta^{\hbar}_M$, and $\delta^{\hbar}_S$ is a transformation given in  (\ref{strans2}) with a parameter $\alpha_S$ constructed again from those of  $\delta^{\hbar}_L$ and $\delta^{\hbar}_M$. Next  $\delta^\hbar_{JP}$ is a transformation again generated by  $\hat\nabla^{\hbar}$-covariantly constant forms collectively denoted by $P$ but now with parameters constructed from those  $\delta^{\hbar}_L$ and $\delta^{\hbar}_M$ and some conserved currents $J$ of the theory. A precise explanation of the structure of $\delta^\hbar_{JP}$ will be given below.   We shall refer to three type of transformations that occur in the right hand side of a commutator as type I, type II and type III, respectively. Of course a typical commutator will close to a linear combination of all 3 types of transformations. The consistency condition in each case will be separately treated.

Type I:  It is clear from (\ref{concon}) that if a commutator closes to a $\delta^{\hbar}_{N}$ type of symmetry, then the consistency condition is
\bea
P(\omega, \Omega)_{\mu\nu[\rho|\sigma|} L^\mu{}_L M^\nu{}_{M]}=0~,
\label{concona}
\eea
as the first term vanishes because $N$ is $\hat\nabla^{\hbar}$ covariantly constant and $\partial_=a_N=0$. Note that the condition is expressed in terms of $L$ and $M$ as $P(\omega, \Omega)$ is first order in $\hbar$.

Type II: On the other hand, if the commutator closes to a $\delta^{\hbar}_S$ symmetry, the consistency condition receives a contribution from the first term in (\ref{concon}). This is because the proof of the  invariance of the classical action for an $S$ type of symmetry involves the Bianchi identity (\ref{bia1}) which gets modified to
\bea
\hat R_{\mu [\nu,\rho\sigma]}=-{1\over3} \hat\nabla_\mu H_{\nu\rho\sigma}-{1\over6} dH_{\mu\nu\rho\sigma}~,
\label{bia2}
\eea
for $dH\not=0$.  This is the case here as $d H^{\hbar}\not=0$. Therefore the consistency condition now reads
\bea
\alpha_S {(-1)^q\over 3( q+1)} P(\omega, \Omega)_{\sigma[\rho\lambda\tau} Q_{Q]}  + a_M a_L P(\omega, \Omega)_{\mu\nu[\rho|\sigma|} L^\mu{}_L M^\nu{}_{M]}=0~,
\label{conconb}
\eea
where $\alpha_S$ is expressed in terms of $a_L$ and $a_M$, and the indices satisfy $\lambda\tau Q=LM$.

Type III:  Finally suppose that a commutator closes to  a $\delta^\hbar_{JP}$ type of symmetry. The type of transformation  that $\delta^\hbar_{JP}$ represents has to be investigated further as a straightforward modification of a transformation generated by a covariantly constant form $P$ by allowing the parameter $a_P$ to depend on $\sigma^=$, $\partial_=a_P\not=0$, is not always\footnote{Though in some cases, $a_P$ depends on the currents of the theory in such a way that $\delta^\hbar_{JP}$ is a symmetry, see $SU$ examples below.} a symmetry of the classical action.  So it is not expected to be a symmetry of the quantum theory. As the commutator of two symmetries $\delta_L$ and $\delta_M$  of a classical action is also a symmetry of the theory, $\delta_{JP}$ has to be a symmetry as well.  It turns out that this is possible if there exist $L'$ and $M'$ $\hat\nabla$-covariantly constant forms that satisfy (\ref{invcon}) such that
 \bea
 \delta_{JP}=(m'+1)c_{L'} J_{L'} \delta_{M'}+ (\ell'+1) c_{M'} J_{M'} \delta_{L'}~,
 \label{typeC}
 \eea
 for some constants $c_{L'}$ and $c_{M'}$ and with parameters related\footnote{As the transformation $\delta_{JP}$ arises in the right hand side of the commutator $[\delta_L, \delta_M]$, the parameters $a_{L'}$ and $a_{M'}$ are expressed in terms of $a_L$ and $a_M$ and so they are related.}  as $(-1)^{(\ell'+1) (m'+1)} c_{L'} a_{M'} =c_{M'} a_{L'}$, where $J_{L'}$ is the current associated to $L'$ as in (\ref{curL}) and similarly for $J_{M'}$. Indeed, one has that
 \bea
  \delta_{JP} S&=&-i\int d^2\sigma d\theta^+ \big( \delta_{JP} X^\mu {\cal S}_\mu\big)
 \cr
 &=&-i \int d^2\sigma d\theta^+  \big(-2 g_{\mu\nu}[
(m'+1) c_{L'} J_{L'} a_{M'} (M')^\mu{}_{M'} D_+X^{M'}
\cr
&&\qquad+
(\ell'+1) c_{M'} J_{M'} a_{L'} (L')^\mu{}_{L'} D_+X^{L'}]  \hat\nabla_= D_+X^\nu\big)
\cr
&=& -i\int d^2\sigma d\theta^+ \big(-2 (-1)^{\ell' m'} c_{M'} a_{L'} \partial_=(J_{M'} J_{L'})\big)=0~,
 \eea
 where we have used the condition on $F$ in (\ref{invcon}) for both the forms $L'$ and $M'$.
 This computation can be repeated in the quantum theory. It turns out that the first term in the consistency condition (\ref{concon}) vanishes after a computation similar to that explained above and so it is required that (\ref{concona}) must be satisfied.

The above treatment of type III commutators can be extended to the case that $\delta_{JP}$ can be written as $\delta_{JP}=c_BJ_A \delta_B+ c_A J_B \delta_A$, where $J_A$ ($J_B$) is a current associated a symmetry $\delta_A$ ($\delta_B$) not necessarily generated by a $\hat\nabla$-covariantly constant form. To indicate how this will work, note that to calculate the current for a symmetry generated by the variation $\delta_B$, one may allow the parameter $a_B$ to dependent on $\sigma^=$.  Then it is known that
\bea
\delta_B S\sim \int d^2\sigma d\theta^+ \partial_=a_B J_B~.
\eea
Using the above formula of calculating a current and after replacing $a_B$  with $J_A a_B$, where now $\partial_=a_B=0$, and similarly for $J_A$,
 one finds that
\bea
\delta_{JP}S\sim  \int d^2\sigma d\theta^+ a_A \partial_=(J_A J_B)=0~,
\label{ABsym}
\eea
 after an appropriate choice of constants $c_A$ and $c_B$ and a relation amongst the parameters $a_A$ and $a_B$.
However if the Bianchi identity (\ref{bia1}) is used to arrange the $\delta_{JP}$ variation of the action as above, then the consistency of anomalies will require the condition (\ref{conconb}) instead of (\ref{concona}).

\section{Anomalies and  holonomy $SU(2)$ backgrounds}

\subsection{Summary of the Geometry}

The spacetime of supersymmetric heterotic backgrounds for which the holonomy of $\hat\nabla$ is included in $SU(2)$ admits six $\hat\nabla$-parallel 1-forms $\bbe^a$, $a=0,\dots, 5$ and three $\hat\nabla$-parallel 2-forms $I_r$ such that
 the Lie bracket algebra of the associated vector fields $\bbe_a$ to $\bbe^a$ is a 6-dimensional Lorentzian Lie algebra with self-dual structure constants $H^a{}_{bc}$.   As a result $\bbe_a$ are Killing vector fields\footnote{So far, we have used  $K$  to denote the Killing vector fields and we shall continue to do so in the analysis of the anomalies that will follow. But here, we stress with $\bbe^a$ that these 1-forms can be used as part of a pseudo-orthonormal (co-)frame on the spacetime.}.    In addition, we have that ${\cal L}_{\bbe_a} H=0$.  Moreover,
 \bea
 i_{\bbe_a} I_r=0~,~~~{\cal L}_{\bbe_a} I_r=0~.
 \eea
 Furthermore, the endomorphisms (vector 1-forms) $I_r$, $g(X, I_r Y)= I_r(X,Y)$, satisfy
 \bea
 I_r I_s=-\delta_{rs} ({\bf 1}-\bbe_a\otimes \bbe^a)+ \epsilon_{rs}{}^t\, I_t~.
 \eea
These backgrounds admit 8 Killing spinors and all these forms arise as Killing spinor bilinears.

The metric and 3-form field strength of the backgrounds   can be written as
\bea
g=\eta_{ab} \bbe^a \bbe^b + \tilde g~,~~~H={1\over3} \eta_{ab} \bbe^a \wedge d\bbe^b+{2\over3} \bbe^a\wedge {\cal F}^b+ \tilde H~,
\eea
where $\eta$ is the Minkowski space metric and  $\tilde g=\delta_{ij} \bbe^i \bbe^j$ with $\bbe^i$  an orthonormal (co)-frame orthogonal to $\bbe^a$.  Moreover, ${\cal F}^a= d\bbe^a-{1\over2} H^a{}_{bc} \bbe^b\wedge \bbe^c={1\over2} H^a{}_{ij} \bbe^i\wedge \bbe^j$ and   $i_{\bbe_a} \tilde H=i_{\bbe_a}{\cal F}^b=0$. Furthermore,
${\cal L}_{\bbe_a} \tilde H=0$ and ${\cal F}^a$ is an (1,1)-form\footnote{Here, we have adopted the terminology of hypercomplex geometry to assign a holomorphic and   anti-holomorphic degree for forms even though $I_r$ is not a hypercomplex structure over the whole spacetime.} with respect to all three endomorphisms $I_r$, i.e. ${\cal F}(I_r X, I_r Y)={\cal F}(X,Y)$ ( no summation over $r$).

The Killing spinor equations also restrict the curvature of connection, $F$, of the gauge sector.  The conditions are that $i_{\bbe_a} F=0$ and  ${ F}(I_r X, I_r Y)={ F}(X,Y)$ ( no summation over $r$). Therefore, $F$ is anti-self dual in the directions orthogonal to the orbits of the isometry group.

The geometry of such spacetime, $M^{10}$, locally can be modelled as that of a principal bundle over and HKT 4-dimensional manifold $N^4$ with metric $\tilde g$ and torsion $\tilde H$, principal bundle connection $\bbe^a$ whose curvature is ${\cal F}$ and fibre a group manifold, $G$, whose (Lorentzian) Lie algebra is $\bR^6$, $\mathfrak{sl}(2,\bR)\oplus \mathfrak{su}(2)$ or $\mathfrak{cw}_6$ with self-dual structure constants. Moreover $N^4$ is conformally hyper-K\"ahler, i.e there is a hyper-K\"ahler metric on $N^4$, $\mathring g$, such that $\tilde g=e^{2\Phi} \mathring g$, where $\Phi$ is the dilaton. The hypercomplex structure on $N^4$ is spanned by the three endomorphisms $\tilde I_r$ and the associated K\"ahler forms $\mathring I_r$ are closed. For more details on the geometry of supersymmetric heterotic backgrounds with $SU(2)$ holonomy, see  \cite{uggp1, uggp2, rev}.  Note that $M^{10}$ may  not be a product $G\times N^4$ either topologically or metrically.  The curvature ${\cal F}$ may not be zero.

 \subsection{$SU(2)$ holonomy symmetries and their commutators} \label{s:comalgsu2}

 The symmetries generated by the form bilinears $\bbe_a$ and $I_r$ are
 \bea
 \delta_K X^\mu= a_{{}_K}^a \bbe_a^\mu~,~~~~\delta_I X^\mu=a_I^r (I_r)^\mu{}_\nu D_+X^\nu~,~~~
 \eea
 with $\Delta_K\psi=\Delta_I\psi=0$.
 For later convenience, we use the pseudo-orthonormal frame $(\bbe^A)=(\bbe^a, \bbe^i)$ and define $\delta X^A=\bbe^A_\mu \delta X^\mu$, and similarly
 $D_+X^A=\bbe^A_\mu D_+ X^\mu$ and $\partial_{\pp}X^A=\bbe^A_\mu \partial_{\pp}X^\mu$. In this notation $\delta_K X^a = a_{{}_K}^a$ and $\delta_I X^i=a_I^r (I_r)^i{}_j D_+X^j$ with all the other components of the variations to vanish.
 For the closure of the algebra of the above transformations, the following symmetry\footnote{Although the variations of the symmetries on the fields are given in frame indices, for the computations of the commutators below it is convenient to re-express them in spacetime indices as those in section \ref{sec:g}.} is also required
 \bea
 &&\delta_C X^a=  \alpha_{{}_C} \hat\nabla_+D_+X^a+\hat\nabla_+(\alpha_{{}_C}  D_+ X^a)~,~~~
 \label{newsym}
 \eea
 with $\delta_C X^i=\Delta_C\psi=0$. The $\delta_C$ symmetry is associated to the quadratic Casimir operator of the Lie algebra of isometries and the conserved current is
 \bea
 C= \eta_{ab} \bbe^a_\mu \bbe^b_\nu D_+X^\mu \hat\nabla_+D_+ X^\nu~.
 \eea

 It is straightforward to verify using the algebraic properties of the form bilinears that
\bea
&& [\delta_K, \delta'_K] X^\mu= a_{{}_K}^a a_{{}_K}'^b [\bbe_a, \bbe_b]^\mu=- a_{{}_K}^a a_{{}_K}'^b H_{ab}{}^c \bbe^\mu_c=\delta_K'' X^\mu~,
\cr
&& [\delta_K, \delta_I] X^\mu= a_{{}_K}^a a_I^r ({\cal L}_{\bbe^a} I_r)^\mu{}_\nu D_+ X^\nu=0~,
\label{KIcom}
\eea
where $(a''_{{}_K})^c=- a_{{}_K}^a a_{{}_K}'^b H_{ab}{}^c$.

It remains to compute the commutator of two symmetries generated by $I_r$. Indeed after some computation, one finds that
\bea
&&[\delta_I, \delta'_I] X^a= \delta^{(1)} X^a+\delta^{(2)} X^a+\delta^{(3)} X^a~,
\cr
&&
[\delta_I, \delta'_I] X^i=\delta^{(1)} X^i+\delta^{(2)} X^i+\delta^{(3)} X^i~,
\label{comm2susysu2}
\eea
where the non-vanishing commutators are
\bea
&&\delta^{(1)} X^a=- a_I'^s  a_I^r \delta_{rs} H^a{}_{ij}  D_+X^i D_+ X^j~,~~~
\cr
&&
\delta^{(2)} X^i=(-a_I'^sD_+a_I^r+a_I^rD_+a_I'^s) \delta_{rs} D_+ X^i- (a_I'^sD_+a_I^r+a_I^rD_+a_I'^s) \epsilon_{rs}{}^t (I_t)^i{}_j D_+X^j~,
\cr
&&
\delta^{(3)} X^i=2i a_I'^s  a_I^r \delta_{rs} \partial_{\pp} X^i~.
\eea
To carry out the above computation, it is helpful to notice that  $N(I_r, I_s)^i{}_{jk}=0$.

It is clear from the above that the right-hand-side of the commutator (\ref{comm2susysu2}) can be rewritten as
\bea
[\delta_I, \delta'_I] =\delta_T+\delta_C+\delta_K+\delta''_{I}~,
\label{Icomsu2}
\eea
where $\alpha_{{}_T}=a_I'^s  a_I^r \delta_{rs}$, $\alpha_{{}_C}=-a_I'^s  a_I^r \delta_{rs}$, $a^a_{{}_K}=a_I'^s  a_I^r \delta_{rs}H^a{}_{bc} J_K^b J_K^c$ and
$a_I''^t=- (a_I'^sD_+a_I^r+a_I^rD_+a_I'^s) \epsilon_{rs}{}^t$. Notice that the parameter of the transformation $\delta_K$ in the right hand side of the commutator is field dependent. In particular it depends  quadratically on the currents $J_K$ associated to isometries. Nevertheless, $\delta_K$ with the above field dependent parameter is symmetry of the theory because it can be rewritten as
\bea
\delta_{\bar H} X^a=a_{{}_{\bar H}} H^a{}_{bc} D_+ X^{bc}~,
\label{hbsym}
\eea
for some parameter $a_{{}_{\bar H}}$, $\partial_=a_{{}_{\bar H}}=0$, generated by the $\hat\nabla$-covariantly constant form
\bea
 \bar H={1\over 3!} H_{abc} \bbe^{abc}~.
 \eea
 The covariant constancy of $\bar H$ is a consequence of the Bianchi identity (\ref{bia1}) and the $SU(2)$ holonomy of $\hat\nabla$. In principle, we could have introduced
 (\ref{hbsym}) as an independent symmetry and compute its commutators. This would have the advantage that the algebra of symmetries of the sigma model would have been a standard Lie algebra instead of a W-type of algebra with current dependent structure constants that emerges in (\ref{Icomsu2}).   We followed this route in the beginning but we decided that it was  more economical for the presentation below not to consider (\ref{hbsym}) as an independent symmetry and express the commutator of two $\delta_I$ transformations as (\ref{Icomsu2}).

 The commutator of two $\delta_I$ transformations on $\psi$ (\ref{psicom}) can be expressed as (\ref{Icomsu2}), where
 \bea
 \Delta_{II}\psi^{\mathrm a}_-=\Delta_T\psi^{\mathrm a}_-=- a_I'^s  a_I^r \delta_{rs} F_{\mu\nu}{}^{\mathrm a}{}_{\mathrm b} \psi_-^{\mathrm b} DX^{\mu\nu}~,
 \eea
 and $\Delta_C\psi=\Delta''_I\psi=0$.

The remaining commutators  are
\bea
[\delta_C, \delta_I]=0~,~~~[\delta_K, \delta_C]=\delta'_K~,~~~[\delta_C, \delta'_C]=\delta''_C+\delta''_K ~,~~~
\label{Icomsu22}
\eea
where $a'^a_K= 2i \alpha_C \partial_{\pp} a_K^a+ D_+\alpha_C D_+ a_{K}^a+2 H^a{}_{bc}  \alpha_CD_+ a_K^b J_K^c$,   $a''_{\bar H}=\alpha''_C=  D_+\alpha'_C D_+\alpha_C +2i (\alpha_C'\partial_{\pp} \alpha_C-\alpha_C\partial_{\pp} \alpha'_C)$ and $a''^a_K=\alpha''_C H^a{}_{bc} J^b_K J^c_K$.
Note again that the parameters of the $\delta'_K$ and $\delta''_K$ transformations are field dependent via the currents of the theory and so the algebra of symmetries is a W-algebra. As we have explained the field dependent part of the $\delta''_K$ transformation is a symmetry of the action as it can be interpreted as a $\delta_{\bar H}$ symmetry. This is also the case for the field dependent part of $\delta_K'$ transformation as
$\delta_{\vec H} X^a= a^b_{\vec H} H^a{}_{bc} D_+ X^c$
is a symmetry of the sigma model action this time generated by the $\hat\nabla$-covariantly constant 2-forms $\vec H_a=-{1\over2} H_{abc}\bbe^{bc}$.  But as we have already mentioned, we have not proceed in this way.  Note also that for all transformations in (\ref{Icomsu22}) $\Delta\psi=0$ and $\Delta_{CI}=\Delta_{KC}=\Delta_{CC}=0$. Below,
 we shall demonstrate the  symmetries generated by $K$ and $C$  are not anomalous as their anomalies can be removed with the addition of a finite local counterterm in the effective action of the theory.

 \subsection{Anomalies and consistency conditions}

 The analysis of the anomalies of these models is similar to that of the standard (2,0)-supersymmetric chiral sigma models in \cite{sen, phgpa}.  However, there are some key differences. This class of sigma models exhibits isometries which are potentially anomalous. In addition $I$ is not a complex structure over the whole spacetime and therefore is not associated with a second supersymmetry -- this is also reflected in the commutator of two transformations generated by $I$ given in (\ref{Icomsu2}). Nevertheless, the analysis presented in \cite{phgpa} for (2,0)-supersymmetric sigma models can be suitably modified to apply to this case as follows.

 So far, we have kept the choice of the connection on $M^{10}$ that contributes in the anomalies (\ref{lan}) and (\ref{Lanom})  arbitrary. From now on, we shall set $\omega=\check\omega$, where $\check\omega$ is the frame connection associated with $\check \nabla$ whose torsion is $-H$. It is well known that if $dH=0$, then $\hat R_{\mu\nu, \rho\sigma}=\check R_{\rho\sigma, \mu\nu}$. As the holonomy of $\hat\nabla$ is in $SU(2)$, one concludes that the curvature 2-form $\check R$ satisfies $i_K \check R=0$ and $\check R(I_rX, I_rY)=\check R(X, Y)$  (no summation over $r=1,2,3$).  As a result $i_K P_4(\check R)=0$ and as a consequence of the gaugino KSE $i_K P_4(F)=0$.  Therefore $i_K P_4(\check R, F)=0$.

 Next notice that $\delta_K$, $\delta_I$, and $\delta_C$  commute with both frame rotations $\delta_\ell$ and $\delta_u$ gauge transformations. As a result, their potential anomaly is given as in (\ref{Lanom}) with $\delta_L$ replaced
 with the symmetry under investigation.

 The commutators of  $\delta_K$ with $\delta_K$,  $\delta_I$ and  $\delta_C$  either vanish or close to a type I and a type III transformation, see (\ref{KIcom}) and (\ref{Icomsu22}). Moreover the type III transformation is a symmetry even with the indicated current  dependent parameters.  Therefore consistency of the anomalies requires the condition (\ref{concona}) which is satisfied as $i_K P_4(\check R, F)=0$.  The same is the case for the commutator   of $\delta_C$ with $\delta_I$ in (\ref{Icomsu22}).

 Next the commutator of two $\delta_C$ transformations in (\ref{Icomsu22}) closes to a type II and a type III transformation generated by  $C$ and $K$, respectively. The latter is a symmetry even with the current dependent  parameters given in (\ref{Icomsu2}). It is straightforward to verify that the consistency condition (\ref{conconb}) is also satisfied because $i_K P_4(\check R, F)=0$.

 The commutators of two $\delta_I$ symmetries, (\ref{Icomsu2}), closes to a type I transformation generated by $I$, two type II transformations generated by $T$ and $C$, respectively, and a type III transformation generated by $K$. The latter is a symmetry with the current dependent parameter indicated. The consistency condition that must be satisfied is (\ref{conconb}) with the first term associated to the transformation generated by $T$. The contribution of $C$ and $K$ transformations vanishes as $i_K P_4(\check R, F)=0$. While in the second term in (\ref{conconb}), one sets $L=I_r$ and $M=I_s$. Adding these substitutions, the consistency condition (\ref{conconb}) required   on $P_4(\check R, F)$ is that it should be a (2,2)-form with respect all three endomorphisms $I_r$. It turns out that this is the case as a consequence of the conditions
$\check R(I_r X, I_r Y)=\check R(X,Y)$ and $F(I_r X, I_r Y)=F(X,Y)$ ( no summation over $r$) of the curvature 2-forms.

Thus, we have demonstrated that  the anomalies of all symmetries are consistent at least at one loop. Repeating the analysis of section \ref{s:anom}, one can argue that the anomalies of all these symmetries cancel at least at one loop provided that the forms which generate the holonomy symmetries are corrected as indicated.  Below, we shall describe the cancellation of   some of these anomalies  with the addition of finite local counterterms in the effective action.

\subsection{Anomaly cancellation and finite local counterterms}\label{s:cansu2}

The consistency and  cancellation of anomalies for holonomy symmetries have already been discussed in section \ref{s:anom}. Here we shall argue that under certain conditions the anomalies of the symmetries generated by $K$,  $C$ and $I$ are cancelled  with the addition of finite local counterterms in the effective action.  The cancellation of the global anomaly requires that $P_4(\check R, F)$ is an exact 4-form. Next $P_4(\check R, F)$ satisfies  $i_K P_4(\check R, F)=0$ and ${\cal L}_KP_4(\check R, F)=0$.  Therefore there is a $\tilde P_4$ on $N^4$ such that $P_4(\check R, F)=\pi^* \tilde P_4$, where $\pi$ is the projection
from the spacetime $M^{10}$ to the orbit space, $N^4$,  of the group of isometries.  As $d\tilde P_4=0$, there is $\tilde Q^0_3$ such that $\tilde P_4=d \tilde Q^0_3$. Therefore, one has
\bea
Q_3^0(\check \omega, \Omega)=\pi^* \tilde Q^0_3+ d W~,
\label{qeo}
\eea
where $W$ is a 2-form on $M^{10}$.
This allows   one to add the finite local counterterm
\bea
\Gamma_{(1)}^{\mathrm{fl}}=-{i\hbar\over 4\pi} \int d^2\sigma d\theta^+\, W_{\mu\nu} D_+X^\mu \partial_=X^\nu~,
\label{flc1}
\eea
in the effective action.
The  addition of this finite local couterterm will reexpress the anomalies of all the symmetries written as (\ref{Lanom})  but now with $Q_3^0(\check \omega, \Omega)$ replaced with $\pi^* \tilde Q^0_3$.  As $i_K \pi^* \tilde Q^0_3=0$, this implies that the anomalies of the symmetries
$\delta_K$ and $\delta_C$ can be removed with this finite local counterterm and therefore these transformations are not anomalous. The same would have been the case, if we had considered the symmetries generated by $\bar H$ and $\vec H$ as independent symmetries.

It remains to investigate the cancellation of the remaining anomalies of the theory. For this notice that the Hodge dual of $\tilde P_4$ on $N^4$ taken with respect to the hyper-K\"ahler metric $\mathring g$, ${\mathring\star}\tilde P_4$, is a scalar. As $\tilde P_4$ is exact, ${\mathring \star}\tilde P_4$ is not harmonic. Therefore, there exists\footnote{This is the case provided that $N^4$ is compact. The same applies for $N^4$ non-compact provided that the operator $\mathring\nabla^2$ has an inverse and  ${\mathring \star}\tilde P_4$ is in the range of the operator. For a non-compact example take $N^4=\bR^4$ with the flat metric and $\tilde P_4$  constructed using anti-self-dual instantons, see \cite{phgpd}.} a function $\tilde f$ on $N^4$ such that ${\mathring\star}\tilde P_4=\mathring \nabla^2 \tilde f$, where $\mathring \nabla$ is the Levi-Civita connection on $N^4$ with respect to  $\mathring g$.  As a result, one can write
\bea
\tilde Q^0_3= -{\mathring \star}d \tilde f+d\tilde X~,
\eea
where $\tilde X$ is a 2-form on $N^4$. Observe that $\tilde Q^0_3$ can also be written as
\bea
\tilde Q^0_3= d_r \mathring Y_r+d\tilde X~,~~~{\mathrm {no~summation ~over}~ r}~,
\eea
where $\mathring Y_r=\mathring I_r \tilde f$ and $\mathring I_r$ is the K\"ahler form of the hyper-K\"ahler metric $\mathring g$ on $N^4$ associated to the complex structure $I_r$, and $d_r=i_{{}_{I_r}} d-di_{{}_{I_r}}$.

Next, adding the finite local counterterm
\bea
\Gamma_{(2)}^{\mathrm {fl}}=-{i\hbar\over 4\pi} \int d^2\sigma d\theta^+\, \big((\pi^*\tilde X)_{\mu\nu}+  f\, (\pi^*\mathring g)_{\mu\nu}\big)\, D_+ X^\mu \partial_=X^\nu~,
\eea
 in the effective action, one can demonstrate that the anomalies, $\Delta(a_I)$, associated to the symmetries generated by $I_r$ endomorphisms cancel as well, where $f=\pi^*\tilde f$.

 The addition of the finite local counterterms  $\Gamma_{(1)}^{\mathrm{fl}}$ and $\Gamma_{(2)}^{\mathrm{fl}}$ can modify the frame rotations and gauge transformations anomalies as
 \bea
 \delta_\ell (\Gamma+\Gamma_{(1)}^{\mathrm{fl}}+\Gamma_{(2)}^{\mathrm{fl}})=\Delta(\ell)+\delta_\ell\Gamma_{(1)}^{\mathrm{fl}}+\delta_\ell\Gamma_{(2)}^{\mathrm{fl}}~,
 \eea
 and similarly for the gauge anomaly $\Delta(u)$. These anomalies can be removed after assigning  an anomalous variation to both the spacetime metric $g$ and $b$ coupling of the sigma model action  (\ref{act1}).

 A refinement of this analysis can be achieved provided we assume that the parameters of the frame rotations and gauge transformations depend only on the coordinates of $N^4$. In such a case notice that $Q_3^0$ can be written as
 \bea
 Q^0_3(\check\omega, \Omega)=\pi^*(\mathring\delta {\mathring \star} \tilde f)+ d\pi^* \tilde X+ dW~,
 \eea
 where $\mathring\delta$ is the adjoint of $d$ on $N^4$ with respect to the hyper-K\"ahler metric. If $\ell$ depends only on the coordinates of $N^4$, then
 \bea
 dQ_2^1(\ell, \check\omega)= \pi^*(\mathring\delta {\mathring \star} \delta_\ell \tilde f)+ d\delta_\ell\pi^* \tilde X+d\delta_\ell W~,
 \eea
 where in the first term in the right hand side we have used that $\ell$ does not depend on the fibre coordinates. This equation
  can be re-arranged as
 \bea
 d(Q_2^1(\ell, \check\omega)-\delta_\ell\pi^* \tilde X-\delta_\ell W)=\pi^*(\mathring\delta {\mathring \star} \delta_\ell \tilde f)~.
 \eea
 As the right hand side vanishes along the orbits of the isometry group, it implies that the same is true for the left hand side of the expression above.
 In addition, the Lie derivative of the left hand side of the above expression vanishes along the directions of the orbits of the isometry group. Therefore
 $d(Q_2^1(\ell, \check\omega)-\delta_\ell\pi^* X-\delta_\ell W)$ is the pull-back of a 3-form on $N^4$. Assuming that it is the pull-back  of an exact 3-form on $N^4$ and using the orthogonality of exact and an co-exact forms on $N^4$, one concludes that
 \bea
 Q_2^1(\ell, \check\omega)=\delta_\ell\pi^* X+\delta_\ell W+ dL~,~~~ \delta_\ell \tilde f=0~,
 \eea
 for some 1-form $L$ on the spacetime and, in particular,
 this  implies that the finite local counterterms $\Gamma_{(1)}^{\mathrm {fl}}$ and $\Gamma_{(2)}^{\mathrm {fl}}$ cancel also the spacetime frame rotation anomaly
 $\Delta(\ell)$.  The terms $dL$  can be treated as representing new anomalies, like  the holomophic anomalies in \cite{sen, phgpa}. Such anomalies can be cancelled by a gauge transformation (\ref{gbtrans}) of the coupling $b$.

\section{Anomalies and $SU(3)$ holonomy  backgrounds}

\subsection{Summary of geometry}

The  spacetime $M^{10}$ of supersymmetric heterotic backgrounds for which the holonomy of $\hat\nabla$  is included in $SU(3)$  admits four $\hat\nabla$-parallel 1-forms $\bbe^a$, $a=0,\dots, 3$, one $\hat\nabla$-parallel 2-form $I$ and a $\hat\nabla$-parallel complex 3-form $L$ such that
 the Lie bracket algebra of the associated vector fields $\bbe_a$ of $\bbe^a$ is a 4-dimensional Lorentzian Lie algebra with structure constants $H^a{}_{bc}$.   As a result $\bbe_a$ are Killing vector fields and $i_{\bbe_a} H=d\bbe^a$, and so  ${\cal L}_{\bbe_a} H=0$ as $dH=0$.  Moreover, one has that
 \bea
 i_{\bbe_a} I=0~,~~~{\cal L}_{\bbe_a} I=0~;~~~i_{\bbe_a} L=0~,~~~{\cal L}_{\bbe_a} L=-{i\over6} \epsilon_a{}^{bcd} H_{bcd} L=-{i\over2} H_{aij} I^{ij}~.
 \label{algprop1}
 \eea
 Notice that if the Lie algebra of the isometry group is not abelian,  $L$ is not invariant under the action of the isometry group.

 Furthermore, the algebraic properties of the $\hat\nabla$ covariantly constant forms include
 \bea
 I^2=- ({\bf 1}-\bbe_a\otimes \bbe^a)~,~~~i_IL=3i L~,
 \label{algprop2}
 \eea
 where $g(X, I Y)= I(X,Y)$.  Therefore, $L$ is a (3,0)-form\footnote{Here, we have adopted the terminology of complex geometry to assign a holomorphic and   anti-holomorphic degree for forms even though $I$ is not a complex structure over the whole spacetime.}  with respect to the endomorphism $I$.

The metric and 3-form field strength of the backgrounds   can be written as
\bea
g=\eta_{ab} \bbe^a \bbe^b + \tilde g~,~~~H={1\over3} \eta_{ab} \bbe^a \wedge d\bbe^b+{2\over3} \bbe^a\wedge {\cal F}^b+ \tilde H~,
\eea
where $\tilde g=\delta_{ij} \bbe^i \bbe^j$,  ${\cal F}^a= d\bbe^a-{1\over2} H^a{}_{bc} \bbe^b\wedge \bbe^c={1\over2} H^a{}_{ij} \bbe^i\wedge \bbe^j$ and $\tilde g(\bbe_a, \cdot)=0$,  $i_{\bbe_a} \tilde H=i_{\bbe_a}{\cal F}^b=0$. Moreover
${\cal L}_{\bbe_a} \tilde H=0$ and ${\cal F}^a$ is an (1,1)-form with respect to the endomorphism $I$, i.e. ${\cal F}(I X, I Y)={\cal F}(X,Y)$. $\tilde H$ is a (1,2)- and (2,1)-form with respect to the endomorphism $I$, i.e.
\bea
H_{ijk}- 3H_{pq [i} I^p{}_j I^q{}_{k]}=0~.~~~
\eea

The Killing spinor equations also restrict the curvature of connection, $F$, of the gauge sector.  The conditions are that $i_{\bbe_a} F=0$ and  ${ F}(I X, I Y)={ F}(X,Y)$  and $i_L F=0$. Therefore, $F$ has non-vanishing components only along  the directions orthogonal to the orbits of the isometry group and it is a (1,1)- and traceless-form with respect to the endomorphism $I$, i.e. $F$ satisfies a generalisation  of the Hermitian-Einstein conditions of an $SU(3)$ instanton.

Locally, the geometry of the spacetime, $M^{10},$  can be modelled as that of a principal bundle over and 6-dimensional KT manifold $N^6$ with metric $\tilde g$ and torsion $\tilde H$, principal bundle connection $\bbe^a$ whose curvature is ${\cal F}$ and fibre a group manifold $G$ whose (Lorentzian) Lie algebra is $\bR^4$, $\bR\oplus \mathfrak{su}(2)$, $\mathfrak{sl}(2,\bR)\oplus \mathfrak{u}(1)$ or $\mathfrak{cw}_4$, for more details see \cite{uggp1, uggp2, rev}. The spacetime $M^{10}$ may not be necessarily of product $G\times N^6$ either topologically or metrically.

 \subsection{$SU(3)$-structure symmetries and their commutators}

 The symmetries generated by the $\hat\nabla$-covariantly constant form bilinears $\bbe_a$, $I$ and $L$ are
 \bea
 &&\delta_K X^\mu= a_{{}_K}^a \bbe_a^\mu~,~~~\delta_I X^\mu=a_I I^\mu{}_\nu D_+X^\nu~,~~~
 \cr
 &&\delta_L X^\mu= a_{{}_L}^r ( L_r)^\mu{}_{\nu_1\nu_2} D_+ X^{\nu_1\nu_2} ~,
 \label{su3sym}
 \eea
 where $L_1=\mathrm{Re}\, L$ and $L_2=\mathrm{Im}\, L$, $r=1,2$. We have normalised the form bilinears such that in holomorphic frame indices $I_{\alpha\bar\beta}=-i\delta_{\alpha\bar\beta}$ and $(L_1)=(\epsilon_{\alpha\beta\gamma}, \epsilon_{\bar\alpha\bar\beta\bar\gamma})$ and
 $(L_2)=(-i\epsilon_{\alpha\beta\gamma}, i\epsilon_{\bar\alpha\bar\beta\bar\gamma})$.

 It is straightforward to verify using the algebraic properties of the form bilinears that
\bea
&& [\delta_K, \delta'_K] = \delta''_K ~,~~~ [\delta_K, \delta_I]=0~,
\label{KKcomsu3}
\eea
where $(a''_{{}_K})^c=- a_{{}_K}^a a_{{}_K}'^b H_{ab}{}^c$. The vanishing of the second commutator is a consequence of ${\cal L}_K I=0$ for heterotic backgrounds with $SU(3)$ holonomy. For all transformations in (\ref{KKcomsu3}) $\Delta\psi=0$ and $\Delta_{KK}=\Delta_{KI}=0$.

The commutator of two symmetries generated by $I$ on $X$ is
\bea
[\delta_I, \delta'_I] =\delta_T+\delta_C+\delta_K~,
\label{Icomsu3}
\eea
where $\delta_C$ is defined as in (\ref{newsym}) after appropriately adapting the formulae for backgrounds with holonomy $SU(3)$, and $\alpha_{{}_T}=a_I'  a_I$, $\alpha_{{}_C}=-a_I'  a_I $ and $a_K^a= a_I'  a_I H^a{}_{bc} J^b_K J^c_K$. The $\delta_K$ transformation in the right hand side of the above commutator could be cast as a $\delta_{\bar H}$ transformation (\ref{hbsym}) leading to field independent structure  constants and so to a Lie algebra  instead of the expected W-algebra structure. But the presentation of the commutator algebra as a W-algebra is far more economical. The commutator of two symmetries generated by $I$ on $\psi$ can also be expressed as in (\ref{Icomsu3}) with the understanding that $\Delta_{II}\psi_-^{\mathrm a}=\Delta_{T}\psi_-^{\mathrm a}= -a'_I a_I F_{\mu\nu}{}^{\mathrm a}{}_{\mathrm b} \psi_-^{\mathrm b} DX^{\mu\nu}$ and $\Delta_C\psi=\Delta_K\psi=0$, where we have used that $F$ is a (1,1)-form with respect to the endomorhism $I$ and $i_KF=0$.

Next,  consider the commutator of symmetries generated by $K$ and $L$.  After using the conditions on the geometry of holonomy $SU(3)$ backgrounds, one finds that the non-vanishing commutators are
\bea
&&[\delta_K, \delta_L] X_i={1\over3} a_{{}_K}^a a_{{}_L}^1 H_{apq} I^{pq} ( L_2)_{ij_1j_2} D_+X^{j_1j_2}
\cr && \qquad\qquad\qquad-
{1\over3} a_{{}_K}^a a_{{}_L}^2 H_{apq} I^{pq}   (L_1)_{ij_1j_2} D_+X^{j_1j_2}~.
\eea
Therefore, the commutator is
\bea
[\delta_K, \delta_L]=\delta_L'~,
\label{KLcomsu3}
\eea
where $a'^r_L=-{1\over3} a_{{}_K}^a \epsilon^r{}_s a_{{}_L}^s H_{apq} I^{pq} $ and $\epsilon$ is the Levi-Civita symbol with $\epsilon^1{}_2=1$. For all transformations in (\ref{KLcomsu3}) $\Delta\psi=0$ and $\Delta_{KL}=0$.

The commutator of the symmetries generated by $I$ and $L$ is
\bea
[\delta_I, \delta_L]= \delta_{a_Ia_{{}_L}}^{(1)}+\delta_{a_Ia_{{}_L}}^{(2)}+\delta_{a_Ia_{{}_L}}^{(3)}~,
\label{ILcomsu3}
\eea
where the non-vanishing variations are
\bea
\delta_{a_Ia_{{}_L}}^{(1)}X_a&=&- a_I a_{{}_L}^r H_{amn} I^m{}_{[i} (L_r)^n{}_{jk]} D_+X^{ijk}
\cr
&=&-{1\over6} a_I a_{{}_L}^r H_{apq} I^{pq} (L_r)_{ijk} D_+X^{ijk}~,
\cr
\delta _{a_Ia_{{}_L}}^{(1)}X_i&=& -3 a_I a_{{}_L}^r  (L_r)^m{}_{[jk} H_{|m|p]a} I^p{}_i  D_+X^{jk}  D_+ X^a
\cr
&=&{ 1\over2} a_I a_{{}_L}^r H_{apq} I^{pq} (L_r)_{ijk} D_+X^{jk}  D_+X^a~,
\cr
\delta _{a_Ia_{{}_L}}^{(2)}X_i&=&(-2  a_{{}_L}^r D_+a_I   I_{mj} (L_r)^m{}_{ik}     +a_I D_+ a_{{}_L}^r   I_{mi} (L_r)^m{}_{jk} ) D_+ X^{jk}~.
\eea
To simplify the commutator, we have used  that $N(I, L_r)^i{}_{jkp}=0$ and
\bea
&&I^m{}_i (L_1)_{mjk}= -(L_2)_{ijk}~,~~~I^m{}_i (L_2)_{mjk}= (L_1)_{ijk}~,
\cr
&&
H_{a[i|m|} (L_1)^m{}_{jk]}={1\over6} H_{apq} I^{pq} (L_2)_{ijk}~,~~~H_{a[i|m|} (L_2)^m{}_{jk]}=-{1\over6} H_{apq} I^{pq} (L_1)_{ijk}~.
\eea
From these, it is clear that the commutator on $X$ is
\bea
[\delta_I, \delta_L]=\delta_L'+\delta_K~,
\label{ILcomsu3}
\eea
where
\bea
a'^r_{L}={ 1\over2} a_I a_{{}_L}^r H_{apq} I^{pq} J^a_K+ 2  a_{{}_L}^s D_+a_I \epsilon^r{}_s-a_I D_+ a_{{}_L}^s \epsilon^r{}_s~,~~~        a^a_K=-{1\over6} a_I a_{{}_L}^r H^a{}_{pq} I^{pq} J_{L_r}~.
\eea
The commutator of these transformations on $\psi$ is as in  (\ref{ILcomsu3}). Note that $\Delta_{IL}=0$ as a consequence of $i_LF=0$ which is consistent with the vanishing of $\Delta\psi$ for all transformations contributing to (\ref{ILcomsu3}).

Therefore the structure constants of the algebra of symmetry transformations depend on the conserved currents $J_K$ and $J_L$ of the theory. This leads   to a W-algebra structure as expected.

It remains to compute the commutator of two transformations generated by $L$.  An outline of this computation is given in appendix B. In particular, one finds that
\bea
[\delta_{L_1}, \delta_{L_2}] = \delta_S+ \delta_I+\delta_K+\delta_C~,
\label{Lcomsu3}
\eea
where the transformation $\delta_S$ is given in (\ref{strans1}) and (\ref{strans2}) for $Q=-I$, i.e.
\bea
S^\mu{}_{\nu\rho\sigma}= -\delta^\mu{}_{[\nu} I_{\rho\sigma]}~,
\label{LLSsu3}
\eea
\bea
\alpha_S=-6a_{{}_L}^1 a_{{}_L}^2~,~~a_{{}_I}=4 a_{{}_L}^1 a_{{}_L}^2C-{4\over3} a_{{}_L}^1 a_{{}_L}^2 H_{abc} J_K^a J_K^b J_K^c -2a_{{}_L}^1 a_{{}_L}^2 J_{I} J^H_{K}~,
\eea
where $J^H_K$ in $a_{{}_I}$ is $J^H_K=H_{aij} I^{ij} D_+ X^a$, and
\bea
a_{{}_C}=-2 a_{{}_L}^1 a_{{}_L}^2 J_I~,~~~a^b_{{}_K}={1\over2} a_{{}_L}^1 a_{{}_L}^2 H^b{}_{ij} I^{ij} J_I^2+2H^b{}_{cd} J_K^c J_K^d a_{{}_L}^1 a_{{}_L}^2 J_I~.
\eea
Notice that the parameters of the transformations that appear in the right hand side of the commutator depend on the currents $J_I$ and $J_K$  of the theory. Similarly, one finds that
\bea
[\delta_{L_1}, \delta'_{L_1}] = \delta_I~,~~~
[\delta_{L_2}, \delta'_{L_2}] = \delta_I~,
\label{LLcomsu3}
\eea
where $a_I=2  (a_{{}_L}^1 D_+a'^1_{{}_L}-a'^1_{{}_L} D_+a_{{}_L}^1) J_I$ and $a_I=-2  (a_{{}_L}^2 D_+a'^2_{{}_L}-a'^2_{{}_L} D_+a_{{}_L}^2) J_I$, respectively.  This summarises the calculation of the commutators of the original symmetries (\ref{su3sym}) of the theory.

The commutators of $\delta_C$ with $\delta_K$ and $\delta_I$  are given as in (\ref{Icomsu22}) for the $SU(2)$ case , i.e. $[\delta_C, \delta_I]=0$ and $[\delta_K, \delta_C]=\delta_k'$. In addition,
\bea
[\delta_L, \delta_C]=\delta'_L+\delta_K~,
\eea
where
\bea
&&a'{}_L^r=\epsilon^r{}_s a^s_L\big(  \alpha_C  D_+J^a_K +{1\over2} D_+ \alpha_C   J_K^a \big) H_{aij} I^{ij} ~,
\cr
&&
a_K^a=H^a{}_{ij} I^{ij} \epsilon^r{}_s\big( {1\over3} a^s_L \alpha_C  D_+ J_{L^r}+{1\over3} D_+a^s_L \alpha_C   J_{L^r}+{1\over 6} a^s_L D_+\alpha_C  J_{L^r}\big)~.
\eea
This concludes the computation of all commutators of the symmetries of sigma models on heterotic backgrounds with holonomy $SU(3)$.


 \subsection{Anomalies and consistency conditions}

 As for $SU(2)$ holonomy backgrounds, we express the anomaly of frame  rotations $\Delta(\ell)$ in terms of the connection $\check \omega$  after possibly adding an appropriate finite local counterterm in the effective action of the theory.  Next, as the commutator of frame rotations $\delta_\ell$ and gauge transformations $\delta_u$ with $\delta_K$,
 $\delta_I$ and $\delta_L$ vanishes, the anomalies of the latter, $\Delta(a_K)$, $\Delta(a_I)$ and $\Delta(a_L)$, respectively, are given as in (\ref{Lanom}) for $L=K, I, L_1, L_2$.  Additional consistency conditions are imposed on all these anomalies that arise from the commutator algebra of $\delta_K$,
 $\delta_I$,  $\delta_{L_1}$  and $\delta_{L_2}$. It is straightforward to see that the commutators of  $\delta_K$ with  $\delta_K$, $\delta_I$,  $\delta_{L_1}$  and $\delta_{L_2}$, eqns (\ref{KKcomsu3}) and (\ref{KLcomsu3}),  either vanish of close to a type I transformation.  So the consistency of the anomalies requires the condition (\ref{concona}). This  is satisfied as  $i_K P_4(\check R, F)=0$.

 The  commutator of two  $\delta_I$ transformations in (\ref{Icomsu3}) closes to two type II transformations generated by $T$ and $C$, respectively, and a type III transformation generated by $K$.  The latter is a symmetry with the indicated current dependence of the parameter.  The consistency condition on the $\Delta(a_I)$ anomaly that arises is given  in (\ref{conconb}) for $S=T$ and $S=C$.  This is satisfied as $i_K P_4(\check R, F)=0$ and  $P_4(\check R, F)$ is (2,2)-form with respect to the endomorphism $I$.

 Next consider the consistency condition on the anomalies that arise from the commutators $[\delta_I, \delta_{L_1}]$  and $[\delta_I, \delta_{L_2}]$ in (\ref{ILcomsu3}). In the former case, the commutator closes in a type I transformation generated by $L_2$ and a type III transformation, (\ref{typeC}),  generated by $L'=K$ and $M'=L_1$. For both cases the consistency condition is given in (\ref{concona}) for $L=I$ and $M=L_1$ which is satisfied as $P_4(\check R, F)$, $I$ and $L_1$ are (2,2)-, (1,1)- and (3,0)+(0,3)-forms with respect to the endomorphism $I$, respectively. The commutator $[\delta_I, \delta_{L_2}]$ can be treated in a similar way.

 The  commutators $[\delta_{L_1}, \delta_{L_1}]$ and $[\delta_{L_2}, \delta_{L_2}]$ in  (\ref{LLcomsu3})  close to a type III transformation, (\ref{typeC}),  generated by $L'=M'=I$.  Therefore the consistency condition on the anomalies is given in (\ref{concona}) for  $L=M=L_1$ and $L=M=L_2$, respectively. This is satisfied as a consequence of the skew-symmetric properties of the above forms in the condition (\ref{concona}).

  The commutator $[\delta_{L_1}, \delta_{L_2}]$ in  (\ref{Lcomsu3}) closes to type II transformations generated by $S$ in (\ref{LLSsu3}) and $C$,  and type III transformations.  The latter are associated to the symmetries generated by the pair of tensors $(A, B)=(I, C)$ and $(L', M')=(I, K)$, see (\ref{ABsym}) and (\ref{typeC}), respectively. After some computation, the consistency condition (\ref{conconb}) can be written as
 \bea
- {2\over3}P_4(\check R, F)_{[\mu_1\mu_2\mu_3|\sigma|} I_{\mu_4\mu_5]}+    P_4(\check R, F)_{\lambda \rho [\mu_1|\sigma|} (L_1)^\lambda{}_{\mu_2\mu_3} (L_2)^\rho{}_{\mu_4\mu_5]}=0~.
 \eea
 This is satisfied for all $P_4(\check R, F)$ which are (2,2)-forms with respect to the endomorphism $I$. This is indeed the case as $\check R$ and $F$ are (1,1)-forms with respect to $I$. However, note that there is no need to use that the connections
 $\check R$ and $F$ satisfy in addition the traceless condition with respect to $I$ as it may have been expected.

 Having demonstrated that all the anomalies are consistent at least at one loop, it is clear that all of them cancel provided one assumes that the
 forms that generate the holonomy symmetries are corrected as indicated in section \ref{s:anom}. As in the $SU(2)$ case, some of these anomalies can also be cancelled with the addition of finite local counterterms in the effective action and this will be described below.

\subsection{Anomaly cancellation and finite local counterterms}

Here, we shall demonstrate that the anomalies of the symmetries generated by $K$, $C$ and $I$ cancel after the addition of appropriate finite local counterterms in the effective action of the theory. To begin, one can repeat the argument deployed for sigma models on the backgrounds with holonomy $SU(2)$ in section \ref{s:cansu2} to argue that $ P_4(\check R, F)$ is the pull-back of a form $\tilde P_4(\check R, F)$ on the orbit space $N^6$ of the isometry group. Then, one can write $Q_3^0(\check\omega, \Omega)$ as in  (\ref{qeo}), i.e. $Q_3^0(\check\omega, \Omega)=\pi^*\tilde Q^0_3+dW$,   where now  $\tilde Q^0_3$ is a 3-form on $N^6$.
 Next, one can add a finite local counterterm constructed from $W$ in the effective action as in (\ref{flc1}) and argue that both anomalies generated by $K$ and $C$ cancel.

It remains to specify the finite local counterms needed to cancel the anomaly associated to the symmetry generated by $I$. This calculation can be presented as in \cite{phgpa} for the cancellation of (2,0) supersymmetry anomaly in sigma models. So we shall be brief.
Viewing $\tilde P_4$ as a 4-form on $N^6$, $\tilde P_4(\check R, F)$ is a (2,2) with respect to  the complex structure $I$ on $N^6$. Appealing to the local triviality of Bott-Chern cohomology, i.e. the local $\partial\bar\partial$-lemma,  there is a (1,1)-form $\tilde Y$  with respect to $I$ on $N^6$ such that
\bea
\tilde P_4(\check R, F)=d d_I \tilde Y~.~~~
\eea
  Therefore $\tilde Q_3^0= d\tilde X+ d_I \tilde Y$ for some 2-form $X$ on $N^6$. Using this, one can construct  a finite local counterterm
\bea
\Gamma^{\mathrm{fl}}=-{i\hbar\over 4\pi} \int d^2\sigma d\theta^+\, ( \pi^*\tilde Z_{\mu\nu}+ \pi^*\tilde X_{\mu\nu}) D_+X^\mu \partial_=X^\nu~,
\label{flsu3}
\eea
that cancels  the $\Delta(a_{{}_I})$ anomaly, where  $\tilde Y(\cdot, \cdot)= \tilde Z(\cdot, I \cdot)$.

This counterterm also cancels the frame rotations and gauge transformation anomalies provided that their parameters depend only on the coordinates of $N^6$. To see this as $\delta_\ell Q_3^0(\check\omega, \Omega)=dQ_2^1(\ell, \check\omega)$, we have that $\delta_\ell \big(Q_3^0(\check\omega, \Omega)-d \delta_\ell W)=\delta_\ell \pi^* \tilde Q_3^0= \pi^*  \delta_\ell \tilde Q_3^0$, where in the last step we have used that $\ell$ depends only on the coordinates of $N^6$. Therefore,
$\pi^*  \delta_\ell \tilde Q_3^0= d Q_2^1(\ell, \check\omega)-d\delta_\ell W$.  Thus, there is a $\tilde Q_2^1(\ell)$ such that $\delta_\ell \tilde Q_3^0= d\tilde Q_2^1(\ell)$.  Using $\tilde Q_3^0= d\tilde X+ d_I \tilde Y$, one finds that
\bea
d(\delta_\ell \tilde X-\tilde Q_2^1(\ell))+d_I \delta_\ell  \tilde Y=0
\label{cancon1}
\eea
which in turn implies that $d d_I(\delta_\ell  \tilde X-\tilde Q_2^1(\ell))=0$ and $dd_I \delta  \tilde Y=0$ on $N^6$. Appealing to the local triviality of Aeppli cohomology, we find that $\delta_\ell \tilde X-\tilde Q_2^1(\ell)=d \tilde A+d_I \tilde B$ and $\delta_\ell  \tilde Y=d \tilde V+ d_I \tilde D$. Substituting this into (\ref{cancon1}) and assuming that $\delta_\ell  \tilde Y$ is (1,1) form with respect to the complex structure $I$, we find that
\bea
\delta_\ell \tilde X=\tilde Q_2^1(\ell)+d\tilde A+d_I \tilde V~,~~~\delta_\ell \tilde Y=d\tilde V+d_I i_I\tilde V~.
\eea
Using these and after varying the finite local counterterm with $\delta_\ell$, one finds that the $\Delta(\ell)$ anomaly cancels. The $d\tilde A$ terms can be absorbed  in a gauge transformation, (\ref{gbtrans}),  of $b$ while the remaining terms can be removed by assigning anomalous variations to both $g$ and $b$. Note that $\tilde V$ terms arise
as a consequence of the remaining holomorphic frame rotations, see \cite{sen, phgpa}. A similar analysis can be performed for the anomaly $\Delta(u)$ of the gauge sector transformations.

\section{Concluding remarks}

We have presented all the commutators of the symmetries, which are generated by the $\hat\nabla$-covariantly constant forms,  of a sigma model with target space supersymmetric heterotic backgrounds with $SU(2)$ and $SU(3)$ holonomy.  We have demonstrated in both cases that the algebra of transformations is a W-algebra and its closure requires additional generators which we describe. We also present the Wess-Zumino consistency conditions of the anomalies of these symmetries arising in quantum theory due to the presence of worldsheet chiral fermions in the sigma model action.  We demonstrate that these anomalies are consistent up to at least one loop in perturbation theory.  We also argue that these anomalies can be cancelled either with the addition of  finite local counterterms followed with suitable anomalous variations of the sigma model couplings or with an appropriate quantum correction of the tensors that generate the associated symmetries  in the quantum theory. The latter is consistent with both the anomaly cancellation mechanism for the spacetime frame rotation and gauge sector anomalies \cite{ewch}  as well as the preservation of the form of the Killing spinor equations of heterotic supergravity up to and including two loops in the sigma model perturbation theory \cite{roo}.

So far,  we have not taken into account the possibility that the  anomalies associated to the holonomy symmetries  receive a contribution from the spacetime frame rotations and gauge transformation invariant terms of the effective action. These can potentially give rise to a $\Delta_{\mathrm {inv}}(a_{{}_L})$ term in (\ref{Lanom}). In what follows, we shall expand on an argument outlined in \cite{phgpw2}, see also \cite{sven} for a detailed analysis on the quantisation of the symmetries generated by forms. To simplify the analysis, suppose  that we are looking at such terms for a sigma model with (1,1) worldsheet supersymmetry with vanishing $b$ coupling, $b=0$, and with target space  a Calabi-Yau manifold $N^6$.  Let us assume that $N^6$ is compact and simply connected. For such models, the connection of the gauge sector is ${\cal D}=\nabla$ and so the symmetry of the gauge sector coincides with the frame rotations of the spacetime. The spacetime frame rotation symmetry of such models is not expected to be anomalous as there is a balance between left-handed and right-handed worldsheet fermions in the sigma model action. Therefore, if there is an anomaly in the symmetries generated by $\nabla$-covariantly constant forms, it will arise from the covariant terms of the effective action.

Sigma models with (1,1) worldsheet supersymmetry, $b=0$ and target space a Calabi-Yau manifold, $N^6$,  are invariant under (2,2) worldsheet supersymmetry transformations generated by the complex structure $I$ on $N^6$. Moreover, they admit an off-shell formulation in terms of (2,2) superfields.  Therefore,  the perturbation theory can be set up in terms of (2,2) superfields.  Such a formulation will require the introduction of complex coordinates on $N^6$ and in such coordinates the (2,2) supersymmetry transformations are linear in the fields. Therefore, one does not expect any corrections to the complex structure of the Calabi-Yau manifold in the quantum theory.
Moreover, it has been argued that any quantum corrections to the Calabi-Yau metric will preserve the K\"ahler property of the metric, see e.g. \cite{mike1, mike2}.  The contribution to the beta function can be expressed as $\beta_{ij}=R_{ij}-(\nabla_i \nabla_j +I^k{}_i I^m{}_j \nabla_k \nabla_m)S$, where $R_{ij}$ is the Ricci tensor which is the one loop contribution and $S$ is a real polynomial constructed from powers of the curvature of $N^6$ that arise from higher order loop corrections.

Suppose now that both the form $L$ and the metric $g$ receive quantum corrections $L^\hbar$ and $g^\hbar$, respectively. Assuming that there is a scheme that the beta function vanishes $\beta=0$ and requiring that the conditions for invariance of the quantum theory under symmetries generated by  $L^\hbar$ retain their classical form,  i.e. $\nabla^\hbar L^\hbar=0$,   one concludes that  the Ricci tensor of $g^\hbar$ vanishes. Therefore, $g^\hbar$ is a Calabi-Yau metric -- in fact, it can be argued that $g=g^\hbar$. As $\beta=0$ as well, one finds that $S$ is the real part of a holomorphic function  and there are no higher order loop corrections to the beta function.  This is a contradiction as there is a nontrivial four loop correction  \cite{ven}.  This indicates that either the condition $\nabla^\hbar L^\hbar=0$ gets modified in the quantum theory and/or there is an anomaly $\Delta_{\mathrm {inv}}(a_{{}_L})\not=0$.

 Indeed, it has been argued in \cite{mike1, mike2} that in the presence of higher order loop contributions to the beta function,  the Killing spinor equation of the effective supergravity theory reduces on $N^6$ to $\nabla_i\eta+{i\over2} I^j{}_i \partial_j S\eta=0$, where $\eta$ is a complex spinor on $N^6$.  As a result, the (3,0)-form Killing spinor bilinear $L^\hbar$ is not covariantly constant but instead it satisfies $\nabla_i^\hbar L^\hbar+i I^m{}_i \partial_m S L^\hbar=0$. The $SU(3)$ structure is preserved but the second term in the previous equation modifies the equations that $L^\hbar$ satisfies in the quantum theory. It is tempting to interpret   the second term of the equation on $L^\hbar$ as a contribution from an anomaly,  $\Delta_{\mathrm {inv}}(a_{{}_L})\sim \int d^2\sigma d^2\theta a_L I^m{}_i \partial_m S L^\hbar_L D_+X^L D_-X^i$, i.e. as it has been done in (\ref{corran}) for the chiral anomaly.  But the anomaly is not consistent. Of course, one can work in a scheme that $g^\hbar=g$ and declare that $L^\hbar=L$ but in such a case there will be a superconformal anomaly as $\beta\not=0$. So at least, if the symmetry generated by $L$ is anomalous, the anomaly will mix with the superconformal anomaly.  A more detailed picture may emerge by investigating further the operator mixing \cite{sven} of the current $J_L$  with other operators at higher loops.

\section*{Acknowledgments}

EPB is supported by the CONACYT, the Mexican Council of Science.




\section*{A: A refinement of the commutator} \label{ap-a}

In this appendix, we shall summarise some key formulae which are useful in the computation of commutators of holonomy symmetries  long directions orthogonal to the orbits of the isometry group  of supersymmetric heterotic backgrounds with compact holonomy group.
On all such backgrounds, there is  a (co-)frame $(\bbe^a, \bbe^i)$ such that the metric and 3-form field strength  can be written as
\bea
g=\eta_{ab} \bbe^a \bbe^b+ \delta_{ij} \bbe^i \bbe^j~,~~H={1\over3!} H_{abc} \bbe^{abc}+{1\over2} H_{aij} \bbe^{aij}+{1\over3!} H_{ijk} \bbe^{ijk}~,
  \eea
  where  $\bbe^a$ is the co-frame dual to the Killing vector fields $\bbe_a$, $\bbe^a(\bbe_b)=\delta^a_b$.  The spacetime is locally a fibration with the fibre directions spanned by the vector fields $\bbe_a$.
Let us consider the symmetries generated by the form bilinears   $L$ such that $i_{\bbe_a} L=0$ and so
\bea
L={1\over(\ell+1)!} L_{i_1\dots i_{\ell+1}}\, \bbe^{i_1i_2\dots i_{\ell+1}}~.
\eea
Suppose that we consider the component of the commutator (\ref{comm}), expressed in terms of $L$,  $M$ and $H$, long the  $\bbe^i$ directions and denote this component with
\bea
[\delta_L, \delta_M]^\perp= \delta{}_{LM}^{\perp(1)}+  \delta_{LM}^{\perp(2)}+  \delta_{LM}^{\perp(3)}~.
 \eea
 This is similar to considering the commutator (\ref{comm}) for a manifold with metric $ g^\perp=\delta_{ij} \bbe^i \bbe^j$ and torsion $ H^\perp={1\over3!} H_{ijk} \bbe^{ijk}$.  As $L= L^\perp$ and $M=M^\perp$, one can demonstrate that  along the $\bbe^i$ directions the conditions
\bea
(L\cdot M)_{i[L_2, jM_2]}=(-1)^{\ell+1}  P^\perp_{ijL_2M_2}+{m\over2} \delta_{i[j}  Q^\perp_{L_2 M_2]}~,
\cr
(L\cdot M)_{[jL_2, |i|M_2]}=(-1)^{\ell} P^\perp_{ijL_2M_2}+{\ell\over2} \delta_{i[j}  Q^\perp_{L_2 M_2]}~,
\cr
(L\cdot M)_{i[L_2, |j|M_2]}+(i\leftrightarrow j)=\delta_{ij}  Q^\perp_{L_2M_2}-(\ell+m-2) \delta_{(i[\ell_2} Q^\perp_{j) L_3 M_2]}~,
\eea
of
\cite{gpph}  hold for all  form Killing spinor bilinears of such supersymmetric backgrounds.  As a result, one can write
\bea
\delta_{LM}^{\perp(1)} X^i+  \delta_{LM}^{\perp(2)} X^i+  \delta_{LM}^{\perp(3)} X^i= \delta_{ P^\perp} X^i+ \delta_{{\mathcal N}^\perp} X^i+ \delta_{ S^\perp} X^i~,
\eea
where $\delta_{ P^\perp}$ is the transformation on $X^i$ generated by the form ${P^\perp}$ and similarly for $\delta_{{\mathcal N}^\perp} X^i$ with
\bea
 {\mathcal N}^\perp_{iLM}=-(\ell+m+1)\Big[ H^\perp_{jk[i} L^j{}_L M^k{}_{M]}+(-1)^\ell {\ell m\over 6}  H^\perp_{[i\ell_1\ell_2} Q^\perp_{L_3 M]}\Big]~.
\eea
Note that ${\mathcal N}^\perp$ is a skew-symmetric modification of the Nijenhuis tensor.
Furthermore, $ S^\perp$ is constructed  from $\tilde Q$ and $\tilde g$ as
\bea
 S^\perp_{i, j_1\dots, j_{\ell+m-1}}= \delta_{i[j_1}  Q^\perp_{j_2\dots j_{\ell+m-1}]}~.
\eea
 If $ H^\perp$ were closed,  the transformations generated by $ {\mathcal N}^\perp$, $ P^\perp$ and $S^\perp$  would have been symmetries of the action (\ref{act1}) with couplings $g=g^\perp$ and $H=H^\perp$.  However in general $ H^\perp$ is not closed.  Nevertheless, the above formulae are useful to compute the component  $\delta_{LM}^{\perp(1)}+  \delta_{LM}^{\perp(2)}+  \delta_{LM}^{\perp(3)}$ of the commutator (\ref{comm}) as a non-vanishing ${\mathcal N}^\perp$, $ P^\perp$ or $ Q^\perp$ indicates the emergence  of new symmetries.

\section*{B: The L-commutator for $SU(3)$ holonomy backgrounds} \label{ap-b}

The calculation of the commutator $[\delta_{L_1}, \delta_{L_2}]$ on $X$ in (\ref{Lcomsu3}) is routine but rather long. So here, we summarise some of the steps in the computation and state some of the formulae we have used. First,  the non-vanishing components of the Nijenhuis tensor  $N(L_1, L_2)$ are
\bea
&&N_{i,jkmn}={2\over3} I_{i[j} H_{kmn]}-2 \delta_{i[j} I^p{}_k H_{mn]p}~,
\cr
&&N_{a,ijk\ell}={1\over2} H_{apq} I^{pq} I_{[ij} I_{k\ell]}-2 H_{a[ij} I_{k\ell]}~,
\cr
&&
N_{i, a jk\ell}=-{1\over2} H_{apq} I^{pq} I_{[ij}I_{k \ell ]}~.
\eea
This may indicate that there is a new symmetry generated by a 5-form constructed from the Nijenhuis tensor  $N(L_1, L_2)$ but this is not the case as the skew-symmetric modification of the Nijenhuis tensor \cite{gpph}, $\tilde {\mathcal N}(L_1, L_2)$, vanishes, see also appendix A. Using,
\bea
(L_1)^{ijk} (L_2)_{mn\ell}=-{9\over4} I^{[i}{}_{[m} \delta^{jk]}_{n\ell]}+{3\over2} I^{[i}{}_m I^j{}_n I^{k]}{}_\ell~,
\eea
one can demonstrate that $\tilde P=0$ while $\tilde Q_{ij}=-I_{ij}$.  Thus one expects a new symmetry of the type (\ref{strans2}) to arise in the right hand side of the commutator $[\delta_{L_1}, \delta_{L_2}]$ generated by $S_{\mu,\nu\rho\sigma}=-g_{\mu[\nu} I_{\rho\sigma]}$.  This is indeed the case as it can be seen in (\ref{Lcomsu3}).

For the calculation of the commutator $[\delta_{L_1}, \delta_{L_2}]$ on $\psi$, notice that $\Delta_{L_1L_2}$ in (\ref{psicom}) and $\Delta_S$ in (\ref{strans2}) for $Q=-I$ are non-zero. For the matching of the left and right hand sides of the commutator on $\psi$, one requires the identity
\bea
F_{\rho\sigma} (L_1)^\rho{}_{[\mu_1\mu_2} (L_2)^\sigma{}_{\mu_3\mu_4]}=2 I_{[\mu_1\mu_2}  F_{\mu_3\mu_4]}~,
\eea
which follows from $i_IF=i_LF=0$. We have also suppressed the gauge indices of the curvature $F$ of the gauge sector.

The calculation of the commutators $[\delta_{L_1}, \delta_{L_1}]$ and $[\delta_{L_2}, \delta_{L_2}]$,  which is much more straightforward,  it is useful to note the identities
\bea
(L_1)^{ijk}(L_1)_{\ell mn}=-(L_2)^{ijk}(L_2)_{\ell mn}={3\over2}\delta^{[i}{}_{[\ell}\delta^{j}{}_{m}\delta^{k]}{}_{n]}-{9\over2}\delta^{[i}{}_{[\ell}\omega^{j}{}_{m}\omega^{k]}{}_{n]}~.
\eea
The result is stated in (\ref{LLcomsu3}). Notice also these commutators on $\psi$ are straightforward as $\Delta_{L_1L_1}=\Delta_{L_2L_2}=0$.

\end{document}